\documentclass[journal]{IEEEtran}
\usepackage{xcolor,soul,framed} 
\colorlet{shadecolor}{yellow}
\usepackage[pdftex]{graphicx}
\graphicspath{{../pdf/}{../jpeg/}}
\DeclareGraphicsExtensions{.pdf,.jpeg,.png}
\usepackage[cmex10]{amsmath}
\usepackage{array}
\usepackage{algorithmic}
\usepackage{algorithm}
\usepackage{amsfonts}
\usepackage{amsmath}
\usepackage{bm}
\usepackage{bbm,dsfont}
\usepackage{booktabs}
\usepackage{blindtext}
\usepackage[colorinlistoftodos]{todonotes}
\usepackage{xcolor}
\usepackage{comment}
\usepackage{amsthm}
\usepackage{cite}
\usepackage{subfigure}
\usepackage{enumerate}
\usepackage{enumitem}
\usepackage{eqparbox}
\usepackage[T1]{fontenc}
\usepackage{footnote}
\usepackage{graphicx}
\usepackage{makecell}
\usepackage{mathtools}
\usepackage{mhchem}
\usepackage{mdwmath}
\usepackage{mdwtab}
\usepackage{multirow}
\usepackage{nomencl}
\usepackage{stfloats}
\usepackage{textcase}
\usepackage{threeparttable}
\usepackage{url}
\usepackage{setspace}

\newtheorem{lemma}{Lemma}
\newtheorem{condition}{Condition}

\begin{document}

\title{\vspace{-0.48cm}\!Two-Stage TSO-DSO Services Provision Framework for Electric Vehicle Coordination}

\IEEEaftertitletext{\vspace{-1.08\baselineskip}}

\author{Yi Wang,~\IEEEmembership{Member,~IEEE,}
        Dawei Qiu,~\IEEEmembership{Member,~IEEE,}
        Fei Teng,~\IEEEmembership{Senior Member,~IEEE,}
        and Goran Strbac,~\IEEEmembership{Member,~IEEE}

\thanks{This work was supported by the UK EPSRC project: `Integrated Development of Low-Carbon Energy Systems (IDLES): A Whole-System Paradigm for Creating a National Strategy' (project code: EP/R045518/1) and the Horizon Europe project: `Reliability, Resilience and Defense technology for the griD' (Grant agreement ID: 101075714).}
}
\markboth{IEEE Trans. Power Syst., Accepted for Publication}%
{Shell \MakeLowercase{\textit{et al.}}: Bare Demo of IEEEtran.cls for IEEE Journals}
\maketitle

\begin{abstract}
High renewable penetration has been witnessed in power systems, resulting in reduced system inertia and increasing requirements for frequency response services. Electric vehicles (EVs), owing to their vehicle-to-grid (V2G) capabilities, can provide cost-effective frequency services for transmission system operators (TSOs). However, EVs that are inherently connected to distribution networks may pose voltage security issues for distribution system operators (DSOs) when supporting TSO frequency. To coordinate both TSO frequency and DSO voltage, this paper proposes a two-stage service provision framework for multi-EVs. At stage one, EVs participate in day-ahead TSO-DSO interactions for frequency reserve schedules; at stage two, EVs make real-time dispatching behaviors in distribution networks for reserve delivery while supporting DSO voltage. Considering the potentially large EV number and environment complexity, a decentralized operation paradigm is introduced for real-time EV dispatches at stage two, while a communication-efficient reinforcement learning (RL) algorithm is proposed to reduce the communication overhead during large-scale multi-agent RL training without compromising policy performance. Case studies are carried out on a 6-bus transmission and 33-bus distribution network as well as a 69-bus distribution network to evaluate the effectiveness and scalability of the proposed method in enabling EVs for frequency service and voltage support.
\end{abstract}
\vspace{-0.2em}
\vspace{-0.98em}

\renewcommand{\nomgroup}[1]{%
\ifthenelse{\equal{#1}{A}}{\item[\emph{A.~Indices~and~Sets}]}{%
\ifthenelse{\equal{#1}{B}}{\item[\emph{B.~Parameters}]}{%
\ifthenelse{\equal{#1}{C}}{\item[\emph{C.~Variables}]}{{}}}}
}
\makenomenclature
\setlength{\nomlabelwidth}{2.05cm} 
\setlength{\nomitemsep}{0.28mm} 
\nomenclature[A1]{$i \in \mathcal{I}$}{Index and set of EV agents}
\nomenclature[A2]{$t \in T$}{Index and set of time steps (15 mins)}

\nomenclature[B01]{$\lambda_{t}^{FR}$}{Shadow prices of FR service ({\pounds}/kWh)}
\nomenclature[B02]{$\lambda_{b,t}^{P},\lambda_{b,t}^{Q}$}{Active and reactive LMPs ({\pounds}/kWh)}
\nomenclature[B03]{$\lambda_{b,t}^{T}$}{LMPs of transmission bus $b$ ({\pounds}/kWh)}
\nomenclature[B04]{$c_{g,t}^{gen}$}{Marginal generation cost of producer $g$ ({\pounds}/kWh)}
\nomenclature[B05]{$x_{bp}$}{Reactance of line $b-p$ (p.u.)}
\nomenclature[B06]{$\underline{\theta},\overline{\theta}$}{Voltage angle limits (rad)}
\nomenclature[B07]{$\underline{P}_{bp},\overline{P}_{bp}$}
{Power flow limits of line $b-p$ (kW)}
\nomenclature[B08]
{$T^{ev}$}{Time constant of EV dynamics (s)}
\nomenclature[B09]{$\underline{E}_{i}^{ev},\overline{E}_{i}^{ev}$}{Energy capacity of EV $i$ (kWh)}
\nomenclature[B10]{$\overline{P}_{i}^{ev}$}{Power capacity of EV $i$ (kW)}
\nomenclature[B11]{$\overline{S}_{i}^{ev}$}{Apparent power capacity of EV $i$ (kVA)}
\nomenclature[B12]{$\eta_{i}^{ev}$}{Charging/discharging efficiency of EV $i$}
\nomenclature[B13]{$f_{0}$}{Nominal grid frequency (Hz)}
\nomenclature[B14]{$\overline{P}^{L}$}{Largest power infeed loss (kW)}
\nomenclature[B15]{$\overline{RoCoF}$}{Rate-of-Change-of-Frequency (Hz/s)}
\nomenclature[B16]{$\Delta\overline{f}$}{Frequency nadir (Hz)}
\nomenclature[C10]{$P^{gen}_{g,t},R^{gen}_{g,t}$}{Generation and reserve of producer $g$ (MW)}
\nomenclature[C11]{$y^{gen}_{g,t}$}{Commitment state of producer $g$ at time $t$}
\nomenclature[C12]{$\theta_{b,t}$}{Voltage angle of bus $b$ (rad)}
\nomenclature[C13]{$P_{i,t}^{c},P_{i,t}^{d}$}{Charging and discharging power of EV $i$ (kW)}
\nomenclature[C14]{$E_{i}^{ev}$}{Battery energy content of EV $i$ (kWh)}
\nomenclature[C15]{$H_{t}$}{System inertia (MW$\cdot$s)}
\nomenclature[C16]{$\tilde{R}^{ev}_{i,t}$}{Day-ahead scheduled FR reserve of EV $i$ (kW)}
\nomenclature[C17]{$R^{ev}_{i,t}$}{Real-time delivered FR service of EV $i$ (kW)}

\printnomenclature
\vspace{-0mm}

\section{Introduction}
\vspace{-0.00em}
\label{sec:I}
Digitization and decentralization are rapidly transforming the power industry and challenging the conventional top-down philosophy of power systems. It can be anticipated that future power systems will be characterized by converter-interfaced renewable energy resources (RESs), e.g., wind and solar \cite{strbac2019cost}. However, due to the lack of system inertia and compensation devices, frequency and voltage instability issues may occur in power system operations, causing security problems. In this context, system operators have to procure an increasing amount of ancillary services for system stability; nevertheless, these traditional ancillary services are typically slow and expensive, which may not be effective for frequency response and voltage regulation of low-inertia power systems \cite{o2022frequency}.

Driven by the inevitable trend of transport electrification, large-scale \textit{electric vehicles} (EVs) will be witnessed in future power systems, with more than 23 million anticipated to be on the road in the UK by 2030. When equipped with vehicle-to-grid (V2G) chargers, large-scale EVs can be regarded as a promising provider of various ancillary services due to their quick response and low investment \cite{o2022frequency}. Specifically, EVs can provide enhanced frequency response (FR) for transmission system operators (TSOs) and voltage regulation (VR) for distribution system operators (DSOs) via their flexibility characteristics. However, both the service provisions of FR and VR require real-time EV dispatches with diverse uncertainties and dynamics \cite{qiu2023reinforcement}. Thus, it is necessary to develop a smart control approach for the multi-EV dispatching problem towards both TSO-FR service and DSO-VR support.

Previous studies have explored the effect of coordinating multi-EVs on TSO-FR service provision, e.g., EV bidding schemes in the FR market \cite{vatandoust2018risk} and FR capacity estimation of EVs \cite{wang2020electric}. However, these papers only consider steady-state FR services that have a slow response time and ignore transient system dynamics. Given the highly scarce inertia of future power systems, post-fault transient system dynamics should be considered for system security \cite{teng2015stochastic}. In this context, there have been studies \cite{thingvad2019value,gao2021optimal,blatiak2022value,o2022frequency} implementing frequency-security constraints in EV-related optimization problems towards secure post-fault frequency evolution. However, the above papers only focus on the day-ahead TSO operation for FR reserve schedules, disregarding how EVs successfully deliver the FR reserve via real-time proactive dispatching behaviors in power distribution networks (PDNs). Additionally, since EVs are firstly connected to PDNs rather than directly providing FR services for TSO, ignoring PDN operations may raise serious security issues, e.g., voltage fluctuations \cite{wang2022evaluation}. In fact, the proliferation of distributed energy resources (DERs) has increased PDN complexity, leading to the need for real-time DSO-VR support. According to control methods, research on coordinating multi-EVs for DSO-VR can be categorized into two groups: 1) \textit{centralized control} \cite{andrianesis2020distribution,pirouzi2020conjugate,pirouzi2017robust} which employs a central controller towards global optimization, while requiring intensive communication resources and raising privacy concerns; 2) \textit{distributed control} \cite{hu2021distributed,wang2018coordinated,andrianesis2021optimal} which eliminates the central controller and enables EVs to develop customized power schedules using local system information while maintaining privacy. However, it can be impractical to adopt model-based optimization approaches for the studied EV dispatching problem considering the large-scale nature of EVs and the necessity for real-time dispatching, since methods in the above papers \cite{andrianesis2020distribution,pirouzi2020conjugate,pirouzi2017robust,hu2021distributed,wang2018coordinated,andrianesis2021optimal} together with \cite{vatandoust2018risk,wang2020electric,thingvad2019value,gao2021optimal,blatiak2022value,o2022frequency} typically demand a thorough comprehension of the power system model, an efficient optimization solver, and accurate uncertainty forecasting.

In this context, deploying a model-free \textit{reinforcement learning} (RL) method \cite{qiu2023reinforcement} for EV service provision problems is appropriate by realizing EVs as agents to learn optimal control policies in a dynamic process without prior knowledge of the power network. In the literature, various single-agent RL (SARL) methods have been applied to the FR \cite{wang2020vehicle,fan2021frequency} or VR \cite{sun2021customized,ding2020optimal,tao2021human,liu2023deep} problems of EVs. However, when addressing the service provision problem involving multi-EVs, the use of SARL methods can result in significant scalability issues due to the curse of dimensionality. In response to this, some research has adopted multi-agent RL (MARL) methods (e.g., multi-agent deep deterministic policy gradient (MADDPG) \cite{wang2022coordinated} and multi-agent actor-critic (MAAC) \cite{hu2022multi}) for DSO-VR problems. Despite these advances, two research gaps remain. On one hand, no work has applied MARL methods to the coordinated FR and VR provision problem of multi-EVs in the context of TSO-DSO interactions. Note that both TSO-FR service and DSO-VR support require real-time EV dispatching, even though the FR reserve is scheduled for EVs by day-ahead TSO operation. On the other hand, existing MARL methods \cite{wang2022coordinated,hu2022multi} adopting centralized training and decentralized execution (CTDE) frameworks normally require intensive information exchange among agents during long-term MARL training and then increase communication burden, which can be impractical. Even though there have been papers (e.g., \cite{zhang2023distributed}) proposing distributed training and distributed execution (DTDE) frameworks for EVs to reduce the amounts of exchanged information during MARL training, the policy update is still based on standard gradient algorithm, requiring communication among agents for every iteration. In this case, the communication overhead during MARL training is still significant, especially given the large-scale nature of EVs.

The above limitations motivate us to develop a novel service provision framework for multi-EVs towards coordinated TSO-FR and DSO-VR support. On one hand, since the FR reserve is first scheduled by day-ahead TSO-DSO interactions and then delivered by real-time EV dispatches, it is necessary to formulate the investigated problem in a two-stage setup to respect operation reality. On the other hand, to reduce the communication burden caused by large-scale MARL training, \textit{Lazily Aggregated Policy Gradient} (LAPG), an extension of the standard policy gradient algorithm, can be employed to adaptively skip policy gradient communications, lowering the communication overhead while still ensuring learning performance \cite{chen2021communication}. In this context, communications among agents can be completely avoided in certain rounds of iterations. Furthermore, since EV agents normally have identical observation, action, and reward functions, it is reasonable to formulate the problem in a distributed framework based on parameter sharing for parallel learning, which allows multi-EVs to learn a common policy, further reducing training time and stabilizing training performance.

The main contributions are summarized as follows:
\begin{enumerate}[label=\arabic*)]
\item Propose a novel two-stage service provision framework considering both day-ahead FR reserve schedules (stage 1) and real-time EV dispatching for coordinated TSO-FR and DSO-VR support (stage 2). In contrast to \cite{vatandoust2018risk,wang2020electric,thingvad2019value,gao2021optimal,blatiak2022value,o2022frequency,andrianesis2020distribution,pirouzi2020conjugate,pirouzi2017robust,hu2021distributed,wang2018coordinated,andrianesis2021optimal,sun2021customized,ding2020optimal,tao2021human,liu2023deep,wang2020vehicle,fan2021frequency,wang2022coordinated,hu2022multi} that focus on either FR or VR problems, this paper models the coordinated FR and VR service provision problem of multi-EVs in the context of TSO-DSO interactions. In contrast to \cite{thingvad2019value,gao2021optimal,blatiak2022value,o2022frequency} that only consider day-ahead TSO-FR schedules, this paper involves both day-ahead FR schedules and real-time FR delivery of large-scale EVs.

\item Formulate the real-time TSO-FR and DSO-VR support problem (stage 2) of multi-EVs as a \textit{Partially Observable Markov Game} (POMG) \cite{qiu2023reinforcement}. In contrast to \cite{vatandoust2018risk,wang2020electric,thingvad2019value,gao2021optimal,blatiak2022value,o2022frequency,andrianesis2020distribution,pirouzi2020conjugate,pirouzi2017robust,hu2021distributed,wang2018coordinated,andrianesis2021optimal} that employ model-based optimization approaches, this paper develops a model-free learning approach without any prior knowledge of the power system model.

\item Present a communication-efficient MARL method to deal with the POMG by i) applying the concept of LAPG, for the first time, to the research area of power system operations, which can reduce the communication burden of large-scale MARL training without degrading policy performance; ii) adopting a fully distributed learning framework based on parameter sharing that allows multi-EVs to learn a common policy with enhanced scalability; iii) deploying a PPO algorithm with an actor-critic architecture to stabilize learning performance for robust MARL policy updates.
\end{enumerate}

The rest of this paper is organized as follows. Section \ref{sec:II} presents the outline of the proposed two-stage service provision framework. Section \ref{sec:III} details the general formulations of multi-EVs for coordinated TSO-FR and DSO-VR provisions, while Section \ref{sec:IV} introduces the communication-efficient MARL method based on PPO. In Section \ref{sec:V}, case studies are carried out and analyzed on two distribution networks. Section \ref{sec:VI} draws the conclusions and future works of this paper.
\vspace{-0.38em}

\section{Outline of the Proposed Two-Stage Framework}
\label{sec:II}
\vspace{-0.00em}
\subsection{Problem Descriptions}
\label{sec:II.A}
This paper focuses on the service provision problem of multi-EVs in the context of TSO-DSO interaction towards coordinated TSO-FR and DSO-VR support. The reason for this combination is that TSO is primarily responsible for maintaining the stability and reliability of the transmission grid. One of the key functions is managing the system frequency, which must be kept within tight bounds around a specified nominal value (e.g., 50 Hz) to avoid grid instability or failure. On the other hand, DSO mainly manages the lower-voltage distribution network that delivers electricity directly to consumers. A critical part of its role is maintaining appropriate voltage levels across the distribution network, which is localized and often requires granular adjustments specific to the distribution level. In detail, considering the day-ahead nature of TSO-FR schedules, the problem is laid out within a two-stage service provision framework, as depicted in Fig. \ref{fig:problem}:

\begin{figure*}[h!] 
\vspace{-0.18em}
\centering  
\includegraphics[width=1\textwidth]{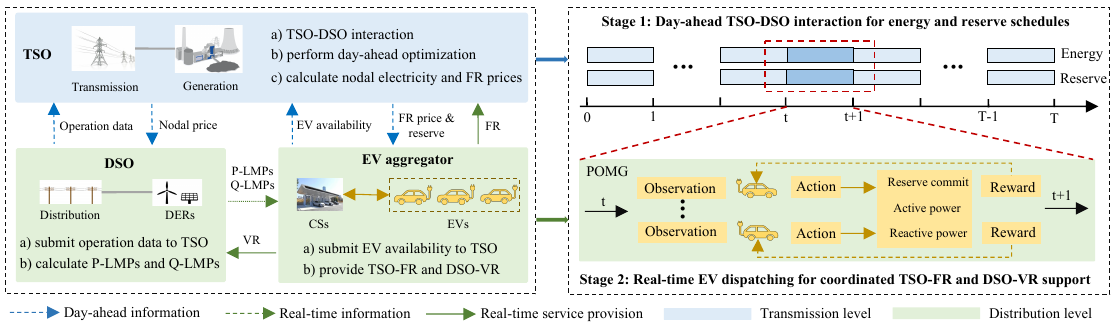}
\vspace{-1.08em}
\caption{Illustration of the proposed two-stage service provision framework of EVs for both TSO-FR and DSO-VR.}
\label{fig:problem}
\vspace{-0.18em}
\end{figure*}

\textbf{Stage 1 - EVs participate in day-ahead TSO-DSO coordination}: a) DSO together with EV aggregator submit operation data and EV availability to TSO; b) TSO performs an optimization for day-ahead energy and reserve schedules; c) TSO calculates nodal electricity prices and FR shadow price signals; d) TSO sends daily locational marginal prices (LMPs) to DSO, while sending FR price signals and reserve schedules to EV aggregator \cite{givisiez2020review}.

Note that there are three types of models that can be used for TSO-DSO interactions, i.e., TSO-managed, TSO-DSO hybrid-managed, and DSO-managed models \cite{givisiez2020review}. This paper adopts the TSO-managed model for TSO-DSO interactions, where TSO coordinates aggregated energy and service providers.

\textbf{Stage 2 - EVs make real-time dispatching behaviors}: a) DSO calculates active and reactive LMP signals (P-LMPs and Q-LMPs) of the PDN and then sends them to the EV aggregator; b) EVs receive FR price signals, FR reserve schedules as well as P-LMPs and Q-LMPs from EV aggregator; c) EVs make real-time dispatching behaviors via charging stations (CSs) in the PDN to deliver FR service, while making active and reactive power scheduling considering DSO-VR support.

To achieve real-time EV dispatching, this paper develops a model-free operating paradigm, formulating the studied problem as a decentralized POMG without time-consuming optimization. In this context, EVs only rely on local information (e.g., observations, actions, and rewards) for real-time decision making, as depicted in Fig. \ref{fig:problem}.

Furthermore, considering the potential information barrier between TSO and DSO, DSO and EV aggregators may not be willing to share all the detailed information with the TSO. To account for this, it can be assumed that only aggregated information at the distribution level is sent to TSO rather than detailed operation data for each DER at stage 1 \cite{wen2023centralized}. On the other hand, TSO only needs to send the LMPs and FR shadow price signals that are specifically for the bus connected with DSO rather than all the price data. In this case, the granularity of the information exchanged between TSO and DSO can be significantly limited and reduced, while customer privacy, especially at the distribution level, can be better protected. It is also worth noting that the main focus of this paper is to investigate how multi-EVs support TSO-FR and DSO-VR via real-time proactive dispatching behaviors, while TSO-DSO interactions serve as the environment for EV dispatches and different types of interaction models can be deployed for simulations.
\vspace{-0.88em}

\subsection{Research Problems and Solutions}
\label{sec:II.B}
In order to achieve coordinated TSO-FR and DSO-VR provisions, the following challenges should be addressed:

1) \textit{How to enable EVs to deliver FR service for TSO and provide VR support for DSO?} To reach this target, appropriate incentive schemes should be incorporated (e.g., FR shadow price signals from TSO and LMPs from DSO), which can effectively promote EVs to deliver TSO-FR service while considering DSO-VR support.

2) \textit{How to solve the second-stage POMG in complex power environment with various uncertainties and dynamics?} To reach this target, a MARL method can be developed, utilizing a data-driven fashion that eliminates the need for precise probability distributions for uncertainties and instead learns state features directly from dataset \cite{qiu2023reinforcement}.

3) \textit{How to reduce communication overhead during large-scale MARL training while ensuring policy performance?} Considering the large number of EV agents, a communication-efficient learning algorithm based on LAPG can be insightful for MARL training, allowing EV agents to skip certain communication rounds without degrading training performance.
\vspace{-0.08em}

\section{General Formulations of multi-EVs for Coordinated TSO-FR and DSO-VR Provisions}
\label{sec:III}
\subsection{Stage 1: Day-Ahead TSO-DSO Interaction}
\label{sec:III.A}
Day-ahead TSO-DSO interaction is solved by the TSO-managed model based on frequency-constrained optimal power flow (FC-OPF) \cite{badesa2022assigning}. After solving the model, daily LMPs $\lambda_{b,t}^{T}$, FR reserve $\tilde{R}^{ev}_{i,t}$ scheduled to EVs, and FR shadow price signals $\lambda_{t}^{FR}$ can be obtained. 

\subsubsection{Network Operation}
\label{sec:III.A.1}
The objective of TSO operation is to minimize the generation cost of producers in the power transmission network (PTN), which is expressed as
\begin{equation}\label{eq:tso obj}
\begin{split}
    \min_{P_{g,t}^{gen}} \sum\nolimits_{g \in \mathcal{G}}\sum\nolimits_{t \in T} c^{gen}_{g,t} P_{g,t}^{gen}
\end{split}
\end{equation}
\vspace{-0.28em}
where 
\begin{equation}\label{eq:energy bal}
    \!\!\Xi^{tso} = \{P_{g,t}^{gen}, P_{g,t}^{res}, y^{gen}_{g,t}, R_{g,t}^{gen},P_{k,t}^{eva}, \tilde{R}^{ev}_{i,t},\theta_{b,t}\}
\end{equation}
\noindent subject to:
\vspace{-0.28em}
\begin{equation}\label{eq:energy bal}
\begin{split}
    \!\! \sum_{g \in \mathcal{TB}_{gen}}\!\!P_{g,t}^{gen} + \!\!\sum_{g \in \mathcal{TB}_{res}} P_{g,t}^{res} =\!\! \sum_{k \in \mathcal{TB}_{eva}} \!\!P_{k,t}^{eva} + \sum_{d \in \mathcal{TB}_{ed}} \!\!P_{d,t}^{ed} \\+ \sum_{(b,p) \in \mathcal{TL}} \theta_{bp,t}/x_{bp}: \lambda_{b,t}^{T}, \forall b \in \mathcal{TB},\forall t \in T,
\end{split}
\end{equation}
\begin{equation}\label{eq:gen energy range}
    y^{gen}_{g,t}\underline{P}_{g}^{gen} \leq P_{g,t}^{gen} \leq y^{gen}_{g,t}\overline{P}_{g}^{gen}, \forall g \in \mathcal{G},\forall t \in T,
\end{equation}
\begin{equation}\label{eq:gen ramp range}
   \!\!\!y^{gen}_{g,t-1}P_{g}^{RD} \!\leq P_{g,t}^{gen} \!- P_{g,t-1}^{gen} \leq P_{g}^{RU}y^{gen}_{g,t}, \forall g \in \mathcal{G},\forall t \in T,
\end{equation}
\begin{equation}\label{eq:cs power}
    P_{k,t}^{eva} \leq \overline{P}^{eva}_{k}, \forall k \in \mathcal{K}.
\end{equation}
\begin{equation}\label{eq:power flow range}
    - \overline{P}_{bp} \leq \theta_{bp,t}/x_{bp} \leq \overline{P}_{bp}, \forall (b,p) \in \mathcal{TL},\forall t \in T,
\end{equation}
\begin{equation}\label{eq:voltage range}
    -\underline{\theta} \leq \theta_{bp,t} \leq \overline{\theta}, \forall b,p \in \mathcal{TB},\forall t \in T,
\end{equation}
where the nodal power balance at bus $b$ refers to \eqref{eq:energy bal}.
$\theta_{bp,t}$ and $x_{bp}$ represent the voltage angle difference between  bus $b$ and bus $p$ as well as the reactance of line $b-p$. The dual variable $\lambda_{b,t}^{T}$ constitutes LMPs of each transmission bus $b$. Sets $\mathcal{TB}_{gen}$, $\mathcal{TB}_{res}$, $\mathcal{TB}_{eva}$, and $\mathcal{TB}_{ed}$ represent the producers, RESs, EV aggregator, and system demand located at bus $b$, respectively. The generation and ramp-up/down limits of producers are modeled in \eqref{eq:gen energy range} and \eqref{eq:gen ramp range}, respectively. $y^{gen}_{g,t}$ indicates the commitment state of producer $g$ at time $t$, where the producer is `on' if $y^{gen}_{g,t}=1$. Equation \eqref{eq:cs power} corresponds to the power demand/supply of EV aggregator $k$ limited by its power capacity $\overline{P}^{eva}_k$. The power flow and nodal angle are limited by \eqref{eq:power flow range} and \eqref{eq:voltage range}, respectively. The voltage angle difference can be limited within the range of $[-\underline{\theta},\overline{\theta}]$ (e.g., $[-\pi/2,\pi/2]$ or $[-\pi/6,\pi/6]$) to reflect a realistic setting \cite{coffrin2015qc}.

\subsubsection{EV-related Constraints}
The power output of EV aggregator $k$ is the aggregated power charging and discharging of all its connected EVs, i.e., $P_{k,t}^{eva}=\sum\nolimits_{i \in \mathcal{I}_{k}}(P_{i,t}^{c} + P_{i,t}^{d})$, where the scheduling behaviors of these EVs are expressed in \eqref{eq:ev model charge}-\eqref{eq:ev storage soc}. $u^{ev}_{i,t}$ indicates the status of EV $i$ ($u^{ev}_{i,t}=1$ if charging; 0, otherwise), while $A_{i,t}$ represents the connecting status of EV $i$ ($A_{i,t}=1$ if connected to the grid; 0, otherwise). $E^{rd}_{i,t}$ is the energy consumption of EV $i$ for traveling at time step $t$.
\begin{equation}\label{eq:ev model charge}
    0 \leq P^{c}_{i,t} \leq u^{ev}_{i,t} A_{i,t} \overline{P}^{ev}_{i}, \forall i \in \mathcal{I}, \forall t \in T,
\end{equation}
\begin{equation}\label{eq:ev model discharge}
    (u^{ev}_{i,t}-1) A_{i,t} \overline{P}^{ev}_{i} \leq P^{d}_{i,t} \leq 0, \forall i \in \mathcal{I}, \forall t \in T,
\end{equation}
\begin{equation}\label{eq:ev storage energy}
    \underline{E}^{ev}_{i} \leq E_{i,t}^{ev} \leq \overline{E}^{ev}_{i}, \forall i \in \mathcal{I}, \forall t \in T,
\end{equation}
\begin{equation}\label{eq:ev storage soc}
\begin{split}
    \!\!E_{i,t+1}^{ev} = 
    \begin{cases}
    E_{i,t}^{ev} + (P^{c}_{i,t}\eta_{i}^{ev} + P^{d}_{i,t}/\eta_{i}^{ev}) \Delta t & \text{if~} A_{i,t} \!=\! 1 \\
    E_{i,t}^{ev} - E^{rd}_{i,t} & \text{if~} A_{i,t} \!=\! 0
    \end{cases}, \\ 
\end{split}
\end{equation}

\subsubsection{Frequency-related Constraints}
\label{sec:III.A.2}
To maintain post-fault frequency security, the frequency-related constraints \eqref{eq:tso inertia}-\eqref{eq:ev pfr_limit_2} are included in the day-ahead TSO operation \cite{badesa2022assigning}:
\begin{equation}\label{eq:tso inertia}
    H_{t}=\sum\nolimits_{g \in \mathcal{G}}H_{g}\overline{P}^{gen}_{g}y^{gen}_{g,t}-H^{L}\overline{P}^{L}, \forall t \in T,
\end{equation}
\begin{equation}\label{eq:tso rocof}
    (\overline{P}^{L}f_0)/(2H_{t}) \leq \overline{RoCoF}, \forall t \in T,
\end{equation}
\begin{equation}\label{eq:tso qss}
    \sum\nolimits_{g \in \mathcal{G}}R^{gen}_{g,t}+\sum\nolimits_{i \in \mathcal{I}}\tilde{R}^{ev}_{i,t} \geq \overline{P}^{L}:\lambda_{t}^{qss}, \forall t \in T,
\end{equation}
\begin{equation}\label{eq:tso nadir}
\begin{split}
    (\frac{H_{t}}{f_0}-\frac{\sum_{i \in \mathcal{I}}\tilde{R}^{ev}_{i,t}T^{ev}}{4\Delta\overline{f}})\frac{\sum_{g \in \mathcal{G}}R^{gen}_{g,t}}{T^{gen}} \geq \\ \frac{(\overline{P}^{L}-\sum_{i \in \mathcal{I}}\tilde{R}^{ev}_{i,t})^2T^{gen}}{4\Delta\overline{f}}:\lambda^{na,1}_{t},\lambda^{na,2}_{t},\mu_{t}^{na}, \forall t \in T,
\end{split}
\end{equation}
\begin{equation}\label{eq:fr_dg_limit}
    0 \leq R^{gen}_{g,t} \leq y^{gen}_{g,t}\overline{P}^{gen}_{g}-P^{gen}_{g,t}, \forall g \in \mathcal{G}, \forall t \in T,
\end{equation}
\begin{equation}\label{eq:ev pfr_limit_2}
    0 \leq \tilde{R}^{ev}_{i,t} \leq \overline{P}^{ev}_{i}+P^{d}_{i,t}+P^{c}_{i,t}, \forall i \in \mathcal{I}, \forall t \in T,
\end{equation}
where the system inertia $H_{t}$ is calculated in \eqref{eq:tso inertia} by aggregating the inertia of producers, except for the largest power infeed loss $\overline{P}^{L}$. The rate-of-change-of-frequency (RoCoF) limit is guaranteed by \eqref{eq:tso rocof}, while the quasi-steady-state (q-s-s) security is assured in \eqref{eq:tso qss}. The frequency nadir is guaranteed by \eqref{eq:tso nadir}, where $\Delta \overline{f}$ is the maximum frequency deviation. $T^{ev},T^{gen}$ are the FR delivery speed of EVs and producers, respectively. The FR quantity of producer $g$ is limited by \eqref{eq:fr_dg_limit}, while the FR reserve $\tilde{R}^{ev}_{i,t}$ scheduled to EV $i$ is limited by \eqref{eq:ev pfr_limit_2}. It is worth noting that this paper focuses on the perspective of normal operation and frequency-constrained scheduling (e.g., how to prepare sufficient FR reserve for the potential power infeed loss) rather than emergency operations (e.g., how to make corrective actions to mitigate the impact of faults when they actually occur), following the same practice in \cite{teng2015stochastic,o2022frequency}.

\subsubsection{FR Price Signals}
\label{sec:III.A.3}
The dual variables of the nadir constraint \eqref{eq:tso nadir} are defined as $\lambda^{na,1}_{t},\lambda^{na,2}_{t},\mu^{na}_{t}$. To comply with dual feasibility, the following constraint is enforced \cite{taylor2015convex}:
\begin{equation}\label{eq:dual fea}
    \left\lVert \begin{bmatrix}
    \lambda^{na,1}_{t} \\
    \lambda^{na,2}_{t} 
    \end{bmatrix} \right\rVert \leq \mu^{na}_{t}, \forall t \in T,
\end{equation}
which can be applied to deduce the shadow price $\lambda_{t}^{FR}$ for FR service from EVs \cite{badesa2022assigning}:
\begin{equation}\label{eq:dual_fr}
\lambda_{t}^{FR}=\frac{(\lambda^{na,1}_{t}-\mu^{na}_{t})T^{ev}}{4\Delta\overline{f}}+\frac{\lambda^{na,2}_{t}}{\sqrt{\Delta\overline{f}}}+\lambda^{qss}_{t}, \forall t \in T.
\end{equation}
Note that this shadow price signal can directly reflect the true value of FR service for frequency security \cite{badesa2022assigning}, which will be used as an incentive signal to promote EVs for real-time FR service provision at stage 2. 
\vspace{-0.48em}

\subsection{Stage 2: Real-time Decentralized EV Dispatching}
\label{sec:III.B}
\vspace{-0.00em}
In this stage, multi-EVs make real-time dispatching behaviors through CSs in the PDN, delivering the TSO-FR reserve scheduled at stage 1 while considering DSO-VR support. In general, formulating this coordinated TSO-FR and DSO-VR service provision problem for multi-EVs faces several challenges. First, it can be difficult to obtain precise mathematical models and technical parameters for EVs in order to formulate the optimization. Second, even though the optimization can be formulated, solving a stochastic optimization problem for large-scale EV dispatching is normally time-consuming. Additionally, uncertainty distributions may not be accurate enough in comparison with real-world data characteristics. Third, it is challenging to generalize a decision-making scheme adapting to system dynamics in a real-time manner, because an optimization must be resolved for every new system condition.

To address the above challenges, the second-stage dispatching problem of multi-EVs is formulated as a decentralized POMG in a dynamic decision-making process, where EVs are realized as agents and can only observe partial information of the power network environment. Generally, POMG can be realized as a 7-tuple $\langle \mathcal{I}, \mathcal{S}, \{\mathcal{O}_{i}\}, \{\mathcal{A}_{i}\}, \{\mathcal{R}_{i}\}, \mathcal{T}, \gamma \rangle$:

\subsubsection{Agent and Environment}
\label{sec:III.B.1}
Each EV $i \in \mathcal{I}$ is regarded an individual agent which controls its own dispatching behaviors. The PDN operating model is specified as the environment; further formulations are provided in Section \ref{sec:III.B.7}.

\subsubsection{Observation}
\label{sec:III.B.2}
The state $s_{t} \in \mathcal{S}$ of environment 
contains individual observations $\{o_{1,t},...,o_{I,t}\}$ and environment information (e.g., the PDN operation model). Specifically, the observation $o_{i,t}$ of EV agent $i$ is defined as
\begin{equation} \label{eq:observation} 
    o_{i,t} = [\lambda_{b,t}^{P}, \lambda_{b,t}^{Q}, \lambda_{t}^{FR}, \tilde{R}_{i,t}^{ev}, E_{i,t}^{ev},] \in \mathcal{O}_{i},
\end{equation}
which consists of three parts: i) the DSO information of CSs' P-LMPs and Q-LMPs $\lambda_{b,t}^{P},\lambda_{b,t}^{Q}$; ii) the TSO information of FR shadow price $\lambda_{t}^{FR}$ and the scheduled reserve $\tilde{R}_{i,t}^{ev}$; and iii) the battery information of its energy content $E_{i,t}^{ev}$. $\lambda_{b,t}^{P},\lambda_{b,t}^{Q}$ can be obtained from the environment presented in Section \ref{sec:III.B.7}.

\subsubsection{Action}
\label{sec:III.B.3}
The action $a_{i,t}$ of EV agent $i$ is defined as
\begin{equation} \label{eq:action}
    a_{i,t}=[a_{i,t}^{fr}, a_{i,t}^{p}, a_{i,t}^{q}] \in \mathcal{A}_{i},
\end{equation}
which comprises three parts: i) the FR commitment action $a_{i,t}^{fr} \in [0,1]$ represents the magnitude of providing FR service for TSO as a percentage of the scheduled FR $\tilde{R}_{i,t}^{ev}$; ii) the active power action $a_{i,t}^{p} \in [-1,1]$ corresponds to the magnitude of active power charging (positive) and discharging (negative) as a percentage of the power capacity $P_{i,t}^{ev} \in [R_{i,t}^{ev}-\overline{P}_{i}^{ev},\overline{P}_{i}^{ev}]$; iii) the reactive power action $a_{i,t}^{q} \in [-1,1]$ corresponds to the magnitude of providing (positive) and consuming (negative) reactive power for DSO as a percentage of the reactive power limits $Q_{i,t}^{ev} \in [-\sqrt{(\overline{S}_{i}^{ev})^{2}-(P_{i,t}^{ev})^{2})},\sqrt{(\overline{S}_{i}^{ev})^{2}-(P_{i,t}^{ev})^{2})}]$.

\subsubsection{State Transition}
\label{sec:III.B.4}
In the proposed real-time multi-EV dispatching problem, $\Delta t$ refers to one time step (e.g., 15 mins). At time step $t$, EV agent $i$ receives its observation $o_{i,t}$ and then executes the action $a_{i,t}$ to the environment, following a policy $\pi_{i}(a_{i,t}|o_{i,t})$. Afterward, the environment transits into the next state $s_{t+1}$ following the transition function $s_{t+1}=\mathcal{T}(s_{t},a_{1:I,t},\chi)$, which is affected by the stochasticity of environment $\chi_{t}$ as well as state $s_{t}$ and actions $a_{1:I,t}$. In the studied POMG, $\chi_{t}=[P^{ed}_{i,t}, Q^{ed}_{i,t}, P^{res}_{i,t}]$ refer to the exogenous states (active demand, reactive demand, and RESs) that exhibit inherent variability.

The transition of the EV energy content $E_{i,t}^{ev}$ in the PDN is an endogenous state and impacted by agent's action $a_{i,t}$, which is limited by its min/max energy capacities $\underline{E}_{i}^{ev},\overline{E}_{i}^{ev}$, power capacity $\overline{P}^{ev}_{i}$, and efficiency $\eta_{i}^{ev}$, depicted as
\begin{equation} \label{eq:tran_meesc} 
    P_{i,t}^{c}=[\textrm{min}(a_{i,t}^{p} \overline{P}_{i}^{ev}, (\overline{E}_{i}^{ev}-E_{i,t}^{ev}) / \eta_{i}^{ev})]^{+},
\end{equation}
\begin{equation} \label{eq:tran_meesd} 
    P_{i,t}^{d}=[\textrm{max}(a_{i,t}^{p} \overline{P}_{i}^{ev}, (\underline{E}_{i}^{ev}-E_{i,t}^{ev}) \eta_{i}^{ev})]^{-},
\end{equation}
where $[\cdot]^{+}$ and $[\cdot]^{-}$ represent $\max\{\cdot,0\}$ and $\min\{\cdot,0\}$, respectively. The state transition of $E_{i,t}^{ev}$ can be realized as \eqref{eq:ev storage soc} with the condition $A_{i,t}=1$. 

\subsubsection{Reward Function}
\label{sec:III.B.5}
Each EV agent $i$ receives its reward $r_{i,t} \in \mathcal{R}_{i}$ at the end of each time step $t$. From the TSO perspective, the objective of EV agent $i$ is to ensure that the day-ahead FR reserve schedules are delivered on time and maximize revenue for FR provision. From the DSO perspective, the objective of EV agent $i$ is to minimize energy charging cost considering VR support via V2G technique. Thus, the reward function can be expressed as
\begin{equation} \label{eq:reward dso} 
\begin{split}
&r_{i,t} = ~r_{i,t}^{tso}+r_{i,t}^{dso}\\
&=\underbrace{\lambda_{i,t}^{FR}R_{i,t}^{ev}-\kappa(\tilde{R}_{i,t}^{ev}-R_{i,t}^{ev})}_{For~TSO}\underbrace{-\lambda_{i,t}^{P}P_{i,t}^{ev}
+\lambda_{i,t}^{Q}Q_{i,t}^{ev}}_{For~DSO},
\end{split}
\end{equation}
where price signals $\lambda_{i,t}^{FR},\lambda_{i,t}^{P},\lambda_{i,t}^{Q}$ constitute the incentive schemes to promote EVs for TSO-FR and DSO-VR service provisions. $\kappa$ is realized as the penalty factor that is used to penalize the violation extent of FR reserve commitment. 

\subsubsection{Objective}
\label{sec:III.B.6}
The above process is simulated every time step in the associated episode, while every EV agent $i$ generates a trajectory containing its observations, actions, and rewards: $\tau_{i} = o_{i,1},a_{i,1},r_{i,1},o_{i,2},...,r_{i,T}$ over $\mathcal{O}_{i} \times \mathcal{A}_{i} \times \mathcal{R}_{i} \rightarrow \mathbb{R}$. Within the POMG, every EV agent $i$ strives for a best policy $\pi_{i}(a_{i,t}|o_{i,t})$ towards discounted reward maximization:
\begin{equation} \label{eq:rl obj} 
    G_{i} = \sum\nolimits_{t \in T} \gamma^{t} r_{i,t},
\end{equation}
where the discount factor is denoted by $\gamma \in [0,1]$.

\subsubsection{Power Network Environment}
\label{sec:III.B.7}
The studied PDN is operated by DSO using a branch flow (BF) algorithm \cite{wang2022coordinated}. Once EVs are plugged into CSs and make scheduling decisions of both active power $P_{i,t}^{ev}=P_{i,t}^{c}+P_{i,t}^{d}$ and reactive power $Q_{i,t}^{ev}$, the following BF algorithm is solved for each time step $t$.
\begin{equation} \label{eq:dso obj} 
    \!\min_{\Xi^{dso}}\!\sum_{g \in \mathcal{DG}}\!\big(c^{P}_{g}P^{dg}_{g,t} + c^{Q}_{g}|Q^{dg}_{g,t}|\big) +\!\!\!\!\sum_{b \in \mathcal{DB}_{tso}}\!\!\!\!\lambda_{b,t}^{T} (P_{b,t}^{gd}+\epsilon^{gd}Q_{b,t}^{gd}),
\end{equation}
\vspace{-0.28em}
\text{where}
\begin{equation}\label{eq:acopf_var}
\begin{split}
    \Xi^{dso} = \{P_{b,t}^{gd},&Q_{b,t}^{gd}, P^{dg}_{g,t}, Q^{dg}_{g,t}, P^{res}_{g,t}, Q^{res}_{g,t}, \\&P_{bp,t}, Q_{bp,t}, \nu_{b,t}, l_{bp,t}\},
\end{split}
\end{equation}
\text{subject to}
\begin{equation}\label{acpowerbalance}
\begin{split}
    \!\!\!\sum_{b \in \mathcal{DB}_{tso}}\!\!P_{b,t}^{gd} \!+\! \sum_{g \in \mathcal{DB}_{dg}}\!\!P^{dg}_{g,t} \!+\! \sum_{g \in \mathcal{DB}_{res}}\!\!P^{res}_{g,t} \!=\! \sum_{d \in \mathcal{DB}_{ed}}\!\!P^{ed}_{d,t}\! +\\\!\sum_{k \in \mathcal{DB}_{ev}}\!\!P_{k,t}^{ev}\!-\!\!\!\!\sum_{(p,b) \in \mathcal{DL}}\!\!P_{pb,t} \!+\!\!\!\! \sum_{(b,p) \in \mathcal{DL}}\!\!P_{bp,t}: \lambda_{b,t}^{P}, \forall b \in \mathcal{DB},
\end{split}
\end{equation}
\begin{equation}\label{rcpowerbalance}
\begin{split}
     \!\!\!\!\sum_{b \in \mathcal{DB}_{tso}}\!\!Q_{b,t}^{gd}+\!\!\!\sum_{g \in \mathcal{DB}_{dg}}\!\!Q^{dg}_{g,t} +\!\! \sum_{g \in \mathcal{DB}_{res}}\!\!Q^{res}_{g,t} \!=\!\!\! \sum_{d \in \mathcal{DB}_{ed}}\!Q^{ed}_{d,t}+\\\!\!\! \sum_{k \in \mathcal{DB}_{ev}}\!\!Q_{k,t}^{ev}-\!\!\!\!\sum_{(p,b) \in \mathcal{DL}}\!\!Q_{pb,t} + \!\!\!\!\sum_{(b,p) \in \mathcal{DL}}\!\!\!Q_{bp,t}: \lambda_{b,t}^{Q},\forall b \in \mathcal{DB},
\end{split}
\end{equation}
\begin{equation}\label{eq:voltage_flow}
\begin{split}
    \nu_{b,t} - \nu_{p,t} = 2 (r_{bp} P_{bp,t} + x_{bp} Q_{bp,t}) \\- (r^2_{bp}+x^2_{bp})l_{bp,t}, ~\forall (b,p) \in \mathcal{DL},
\end{split}
\end{equation}
\begin{equation}\label{thermal}
    P_{bp,t}^2 + Q_{bp,t}^2 \leq l_{bp,t} \nu_{b,t}, \forall (b,p) \in \mathcal{DL},
\end{equation}
\begin{equation}\label{vol}
    \underline{\nu} \leq \nu_{b,t} \leq \overline{\nu}, \forall b \in \mathcal{DB},
\end{equation}
\begin{equation}\label{ampacity}
    l_{bp,t} \leq \overline{l}_{bp}, \forall (b,p) \in \mathcal{DL},
\end{equation}
\begin{equation}\label{acpowerdg}
    \underline{P}^{dg}_{g} \leq P^{dg}_{g,t} \leq \overline{P}^{dg}_{g}, \forall g \in \mathcal{DG},
\end{equation}
\begin{equation}\label{rcpowerdg}
    (P^{dg}_{g,t})^2 + (Q^{dg}_{g,t})^2 \leq (\overline{S}^{dg}_{g})^2, \forall g \in \mathcal{DG},
\end{equation}
\begin{equation}\label{pf_dg}
    |Q^{dg}_{g,t}| \leq P^{dg}_{g,t} \tan(\cos^{-1}\delta^{dg}_{g}), \forall g \in \mathcal{DG},
\end{equation}
where the objective function \eqref{eq:dso obj} is to minimize the operation costs of active and reactive power of distributed generators (DGs) as well as the energy procurement from TSO at nodal price $\lambda_{b,t}^{T}$. Note that the nodal price $\lambda_{b,t}^{T}$ also represents the LMP at transmission bus $b$ of TSO, reflecting the TSO-DSO interaction process at stage 1. This setting is different with \cite{wang2022coordinated}, which ignores TSO operations and assumes DSO grid prices as input data. The nodal active and reactive power balances at bus $b$ are limited by \eqref{acpowerbalance}-\eqref{rcpowerbalance}, where their dual variables $\lambda_{b,t}^{P}$ and $\lambda_{b,t}^{Q}$ correspond to P-LMPs and Q-LMPs, respectively. Sets $\mathcal{DB}_{tso}$, $\mathcal{DB}_{dg}$, $\mathcal{DB}_{res}$, $\mathcal{DB}_{ed}$, and $\mathcal{DB}_{ev}$ correspond to the TSO-DSO interaction, DGs, RESs, local demand, and EVs located at bus $b$, respectively. Constraints \eqref{eq:voltage_flow}-\eqref{ampacity} model the limits and relationships of squared voltage $\nu_{b,t}$, current $l_{bp,t}$, and line flow $P_{bp,t}/Q_{bp,t}$. Constraints \eqref{acpowerdg}-\eqref{pf_dg} capture the operating region of DGs.

\section{Problem Solving Procedure}
\vspace{-0.00em}
\label{sec:IV}
The day-ahead TSO operation at stage 1 (Section \ref{sec:III.A}) can be solved by commercial software such as Gurobi, while the POMG proposed at stage 2 (Section \ref{sec:III.B}) requires real-time and decentralized decision-making of multi-EVs. Since our primary focus is on stage 2, this section focuses on outlining technique details of the proposed MARL method for stage 2.

\begin{figure}[t!]
\centering
\includegraphics[width=0.48\textwidth]{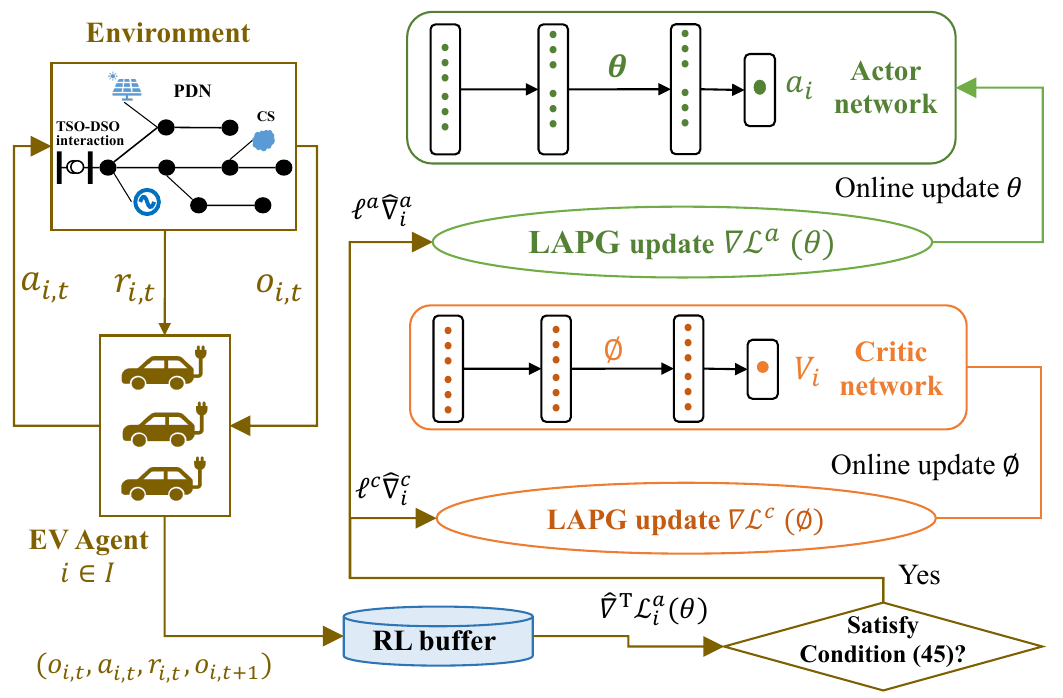}
\vspace{-0.18em}
\caption{Architecture of the proposed communication-efficient learning method in solving the POMG at stage 2.}
\label{fig:la_dppo}
\vspace{-0.38em}
\end{figure}

In detail, a communication-efficient learning method called LA-DPPO is developed for resolving the POMG at stage 2, where the general architecture is depicted in Fig. \ref{fig:la_dppo}. Specifically, LA-DPPO offers three insightful and critical techniques: 1) introducing a distributed framework based on parameter sharing for the parallel learning of a common policy to improve policy scalability and stability; 2) developing a LAPG-based learning algorithm for communication-efficient learning without compromising training performance; 3) utilizing a PPO algorithm with an actor-critic architecture \cite{wang2023coordinating} for stable MARL policy updates because of its stable learning performance and robust hyperparameter tuning.
\vspace{-0.88em}

\subsection{Distributed Learning Framework}
\vspace{-0.00em}
\label{sec:V.A}
Since EV agents have identical features of observation, action, state transition, and reward functions, as defined in Section \ref{sec:III.B}, it is reasonable to formulate the proposed problem in a distributed learning framework based on parameter sharing with a parallel RL setting, which can significantly reduce training time and stabilize training process \cite{chen2021communication,terry2020parameter}. In this setting, multi-EVs learn a common policy $\pi_{\bm{\theta}}$ parameterized by $\bm{\theta}$ for different instances, where rewards and observations can be different for EV agents.

In general, a distributed RL framework includes a central controller and multiple learners (EV agents), where EV agents learn their own behaviors via common policy $\pi_{\bm{\theta}}$. To find the optimal policy $\pi_{\bm{\theta}}$, the objective is defined as the maximization of the advantage function aggregated over all EV agents:
\begin{equation} \label{eq:drl_obj}
    \max_{\bm{\theta}} \sum_{i \in \mathcal{I}}\mathcal{L}_{i}(\bm{\theta})=\sum_{i \in \mathcal{I}}\mathbb{E}_{t}\big[\log \pi_{\bm{\theta}}(a_{i,\tau}|o_{i,\tau})\hat{\Lambda}_{i,t}\big],
\end{equation}
where $\hat{\Lambda}_{i,t}$ is the approximated advantage function of EV agent $i$ at time step $t$, following the common policy $\pi_{\bm{\theta}}$.

Using the \textit{log-trick}, the approximation of the policy gradient for each EV agent $i$ is expressed as
\begin{equation} \label{eq:pg_appro}
\begin{split}
    \nabla_{\bm{\theta}}\mathcal{L}_{i}(\bm{\theta})=\mathbb{E}_{t}\big[\nabla_{\bm{\theta}} &\log \pi_{\bm{\theta}}(a_{i,t}|o_{i,t})\hat{\Lambda}_{i,t}\big].
\end{split}
\end{equation}

Specifically, at iteration $k$, the aggregator broadcasts the current policy $\pi_{\bm{\theta}^{k}}$ to all EV agents; each agent $i \in \mathcal{I}$ computes an approximate policy gradient $\hat{\nabla}_{\bm{\theta}}^{N,T}\mathcal{L}_{i}(\bm{\theta}^{k})$ and uploads it to the aggregator for policy update
\begin{equation} \label{eq:log_loss_each}
\begin{split}
    \bm{\theta}^{k+1}&=\bm{\theta}^{k}+\alpha \hat{\nabla}^{k}\\
&=\bm{\theta}^{k}+\alpha\sum\nolimits_{i \in \mathcal{I}}\hat{\nabla}_{\bm{\theta}}^{N,T}\mathcal{L}_{i}(\bm{\theta}^{k}),
\end{split}
\end{equation}
where $\alpha$ and $N$ are the learning rate and the number of batch trajectories, respectively.

\subsection{Lazily Aggregated Policy Gradient}
\label{sec:V.B}
To implement the policy gradient update in \eqref{eq:log_loss_each}, the aggregator has to communicate with all EV agents to obtain $\{\hat{\nabla}_{\bm{\theta}}^{N,T}\mathcal{L}_{i}(\bm{\theta}^{k})\}$, which requires intensive communication resources, especially when large-scale EVs are integrated. To mitigate this issue, this paper introduces LAPG in the policy gradient update, which can reduce the communication overhead by skipping communication at certain rounds \cite{chen2021communication}.

Instead of requesting policy gradients from every EV agent, LAPG only requires the new gradients from part of the agents $\mathcal{\tilde{I}}^{k} \in \mathcal{I}$, while reusing the old aggregated policy gradient $\hat{\nabla}^{k-1}$ in iteration $k-1$. The rule of LAPG update is expressed as 
\begin{equation} \label{eq:lapg_update}
\begin{split}
    \bm{\theta}^{k+1}=\bm{\theta}^{k}+\alpha \hat{\nabla}^{k-1}+\alpha \sum\nolimits_{i\in \mathcal{\tilde{I}}^{k}}\ell\hat{\nabla}_{i}^{k},\\
    \ell\hat{\nabla}_{i}^{k} = \hat{\nabla}_{\bm{\theta}}^{N,T}\mathcal{L}_{i}(\bm{\theta}^{k})-\hat{\nabla}_{\bm{\theta}}^{N,T}\mathcal{L}_{i}(\bm{\theta}^{k-1}),
\end{split}
\end{equation}
where $\ell\hat{\nabla}_{i}^{k}$ refers to the innovation between two policy gradient evaluations of $\hat{\nabla}_{\bm{\theta}}^{N,T}\mathcal{L}_{i}(\bm{\theta}^{k})$ at the current policy parameter $\bm{\theta}^{k}$ and the old one $\bm{\theta}^{k-1}$. In this way, if the aggregator stores the previous $\hat{\nabla}^{k-1}$, EV agents in $\mathcal{\tilde{I}}^{k}$ only need to upload the policy gradient innovation $\ell\hat{\nabla}_{i}^{k}$. 

It is worth noting that the communication intensity is directly linked to the number of EV agents in $\mathcal{\tilde{I}}^{k}$. More specifically, removing agents from $\mathcal{\tilde{I}}^{k}$ can significantly reduce the required communication rounds for each iteration, while it may also lead to many more iterations for algorithm convergence due to the inefficient policy gradient update, eventually resulting in even more total communication rounds. To address this issue and achieve communication-efficient learning, two lemmas and the selection condition of LAPG update that can guide the communication selection according to learners’ optimization progress are introduced as follows \cite{chen2021communication}: 

\begin{lemma}[\textbf{LAPG Ascent}] \label{lemma_1}
Assume $\mathcal{L}(\bm{\theta})=\sum_{i \in \mathcal{I}}\mathcal{L}_{i}(\bm{\theta})$ is L-smooth, and $\bm{\theta}^{k+1}$ is generated by running one-step LAPG iteration given $\bm{\theta}^{k}$. If the learning rate is selected as $\alpha \leq 1/L$, then the objective values satisfy
\begin{equation} \label{eq:lemma_1}
\begin{split}
\!\!\mathcal{L}(\bm{\theta}^{k})-&\mathcal{L}(\bm{\theta}^{k+1}) \leq \\
-\frac{\alpha}{2}||\nabla_{\bm{\theta}} \mathcal{L}(\bm{\theta}^{k})||^{2}&+\frac{3\alpha}{2}||\nabla_{\bm{\theta}}^{T}\mathcal{L}(\bm{\theta}^{k})-\nabla_{\bm{\theta}}\mathcal{L}(\bm{\theta}^{k})||^{2} \\\!\!+\frac{3\alpha}{2}||\sum_{i \in \mathcal{I}/\mathcal{\tilde{I}}^{k}}\ell\hat{\nabla}^{k}_{i}||^2
&+\frac{3\alpha}{2}||\hat{\nabla}_{\theta}^{N,T}\mathcal{L}(\bm{\theta}^{k})-\nabla_{\bm{\theta}}^{T}\mathcal{L}(\bm{\theta}^{k})||^2\\
&+(\frac{L}{2}-\frac{1}{2\alpha})||\bm{\theta}^{k+1}-\bm{\theta}^{k}||^2,
\end{split}
\end{equation}
where $\mathcal{L}(\bm{\theta}^{k})$ can be approximated by 
\begin{equation} \label{eq:xi_cal}
\sum\nolimits_{d=1}^{D}\frac{\xi}{\alpha^2}||\bm{\theta}^{k+1-d}-\bm{\theta}^{k-d}||^2,
\end{equation}
where $\xi$ is a constant and $D$ is a pre-defined interval length.
\end{lemma}

\begin{lemma}[\textbf{Policy Gradient Concentration}]
Given $K$ and $\ell \in (0,1)$, there exists a constant $U_{i}$ with at least $1-\ell/K$ probability for any $\bm{\theta}$ that satisfies
\begin{equation}\label{eq:lemma_2}
||\hat{\nabla}^{N,T}\mathcal{L}(\bm{\theta}^{k})-\nabla^{T}\mathcal{L}(\bm{\theta}^{k})||^{2} \leq \frac{2\log(2K/\ell)U^2_{i}}{N},
\end{equation}
where $\frac{2\log(2K/\ell)U^2_{i}}{N}$ can be denoted as $\sigma^2$, representing the variance bound on the gradient estimation error \cite{chen2021communication}.
\end{lemma}

\begin{condition}[\textbf{Selection Condition}]
Following the above Lemmas 1 and 2, EV agent $i$ will be included in $\mathcal{\hat{I}}$ for policy gradient update, only if its current policy gradient has enough innovation $\ell\hat{\nabla}^{k}_{i}$. The selection condition can be expressed as
\begin{equation}\label{eq:condition}
\begin{split}
||\ell\hat{\nabla}^{k}_{i}||^2 \geq \frac{\xi}{\alpha^2|\mathcal{I}|^2}\sum_{d=1}^{D}||\bm{\theta}^{k+1-d}-\bm{\theta}^{k-d}||^2+6\sigma^2,
\end{split}
\end{equation}
which is determined by the magnitude of parameter updates and the stochastic variability of gradient estimates.
\end{condition}

As a result, LAPG can be implemented by the policy gradient update rule in \eqref{eq:lapg_update} and the selection condition in \eqref{eq:condition}, which has the same complexity as vanilla policy gradient methods. The only additional complexity comes from storing the policy gradient $\hat{\nabla}_{\bm{\theta}}^{N,T}\mathcal{L}_{i}(\bm{\theta}^{k})$ that is most recently uploaded and checking the selection condition \eqref{eq:condition}. In this context, EV agents in $\mathcal{I}/\mathcal{\tilde{I}}$ do not need to upload gradient calculations, cutting down unnecessary communication rounds.
\vspace{-1.10em}

\subsection{Proximal Policy Optimization}
\vspace{-0.10em}
\label{sec:V.C}
PPO has been used in a variety of control problems as a policy gradient method \cite{wang2023coordinating}. In general, PPO is distinguished by an actor-critic structure which can handle high-dimensional continuous spaces. In order to respect the action spaces in \eqref{eq:action}, a stochastic policy $\pi_{\bm{\theta}}(a|o)$ following Gaussian distributions is created for the actor network parameterized by $\bm{\theta}$, which samples action $a_{i,t}$ conditioned on observation $o_{i,t}$ for power dispatches of EV agent $i$ and outputs related mean $\mu$ and standard deviation $\varsigma$. To maximize the clipped surrogate, the stochastic policy $\pi_{\bm{\theta}}(a|o)$ is updated via policy gradient:
\vspace{-0.18em}
\begin{equation} \label{eq:mappo clip policy}
    L_{i,t}(\bm{\theta}) = \mathbb{E}_{t} \big[\min(\zeta_{i,t}\hat{\Lambda}_{i,t}, \text{clip}(\zeta_{i,t},1-\epsilon,1+\epsilon)\hat{\Lambda}_{i,t} ) \big],
\end{equation}
where the policy gradient is denoted by $\zeta_{i,t}\hat{\Lambda}_{i,t}$, and the probability ratio clipped by $\text{clip}(\cdot)$ is denoted by $\zeta_{i,t}$. $\epsilon \in [0,1]$ is a hyperparameter used to limit the gradient update of current policy from the previous version, in a case that the probability ratio $\zeta_{i,t}$ is outside of the range $[1-\epsilon,1+\epsilon]$. 

In particular, the probability proportion $\zeta_{i,t}$ in the clipped policy \eqref{eq:mappo clip policy} is expressed as
\vspace{-0.68em}
\begin{equation} \label{eq:mappo probability}
    \zeta_{i,t} = \frac{\pi_{\bm{\theta}}(a_{i,t}|o_{i,t})}{\pi_{\bm{\theta}^{\text{old}}}(a_{i,t}|o_{i,t})},
\end{equation}
where $\pi_{\bm{\theta}}(a_{i,t}|o_{i,t})$ and $\pi_{\bm{\theta}^{\text{old}}}(a_{i,t}|o_{i,t})$ stand for current policy and the previous version, respectively. Furthermore, the generalized advantage function $\hat{\Lambda}_{i,t}$ can be written as
\vspace{-0.28em}
\begin{equation} \label{eq:mappo advantage}
    \hat{\Lambda}_{i,t} = \delta_{i,t} + (\gamma\lambda) \delta_{i,t+1} + \dots + (\gamma\lambda)^{T-t+1} \delta_{i,T-1},
\end{equation}
\begin{equation} \label{eq:mappo value function}
    \delta_{i,t} = r_{i,t} + \gamma V_{\bm{\phi}}(o_{i,t+1}) - V_{\bm{\phi}}(o_{i,t}),
\end{equation}
where $V_{\bm{\phi}}(o)$ refers to the state-value function estimated via a critic network parameterized by $\bm{\phi}$. $\gamma \in [0,1]$ and $\lambda \in [0,1]$.
\vspace{-0.68em}

\subsection{Training Procedure}
\label{sec:V.D}
With respect to the training process of LA-DPPO, every agent possesses a common policy $\pi_{\bm{\theta}}(a|o)$ interacting with the environment across $T$ time steps. As a result, a trajectory $\tau_{i}$ is gathered to calculate the discounted reward-to-go $\hat{R}_{i,t} = \sum_{h=t}^{T} \gamma^{h-t} r_{h}$. Using the trajectory $\tau_{i}$, the advantage function $\hat{\Lambda}_{i,t}$ can be estimated via the state-value function $V_{\bm{\phi}}(o_{i,t})$ at each time step $t$. After a batch of trajectories ($N$) are gathered from the buffer $\mathcal{J}_{i} = \{n\} \sim \mathcal{F}_{i}$, the actor network is trained by maximizing the following function:
\vspace{-0.18em}
\begin{equation} \label{eq:actor train}
    \mathcal{L}_{i}^{a}(\bm{\theta}) = \frac{1}{N} \sum\nolimits_{n=1}^{N} \sum\nolimits_{t=0}^{T} L^{n}_{i,t}(\bm{\theta}).
\end{equation}

Consequently, the following loss function of mean-squared error is minimized to train the PPO critic network:
\vspace{-0.28em}
\begin{equation} \label{eq:critic train}
    \mathcal{L}_{i}^{c}(\bm{\phi}) = \frac{1}{N} \sum\nolimits_{n=1}^{N} \sum\nolimits_{t=0}^{T}\big(V_{\bm{\phi}}(o^{n}_{i,t}) - \hat{R}^{n}_{i,t} \big)^{2}.
\end{equation}
\vspace{-0.28em}

Finally, the update rules of actor and critic networks can be expressed as
\vspace{-0.28em}
\begin{equation} \label{eq:actor weight}
    \bm{\theta}^{k+1} = \bm{\theta}^{k} + \alpha^{\bm{\theta}} \nabla^{k-1}_{\bm{\theta}}\mathcal{L}^{a}(\bm{\theta})+\alpha^{\bm{\theta}} \sum\nolimits_{i\in \mathcal{\tilde{I}}^{k}}\ell^{a}\hat{\nabla}_{i}^{a},
\end{equation}
\begin{equation} \label{eq:critic weight}
    \bm{\phi}^{k+1} = \bm{\phi}^{k} - \alpha^{\bm{\phi}} \nabla^{k-1}_{\bm{\phi}}\mathcal{L}^{c}(\bm{\phi})-\alpha^{\bm{\phi}} \sum\nolimits_{i\in \mathcal{\tilde{I}}^{k}}\ell^{c}\hat{\nabla}_{i}^{c},
\end{equation}
where $\alpha^{\bm{\theta}},\alpha^{\bm{\phi}}$ correspond to the learning rates of actor and critic networks, respectively. $\ell^{a}\hat{\nabla}_{i}^{a}$ and $\ell^{c}\hat{\nabla}_{i}^{c}$ denote the innovations between two evaluations of $\nabla_{\bm{\theta}}\mathcal{L}_{i}^{a}$ and $\nabla_{\bm{\phi}}\mathcal{L}_{i}^{c}$ from iteration $k-1$ to $k$, respectively. Finally, the pseudo-code of LA-DPPO can be shown below:
\begin{algorithm}
\setstretch{1.00}
\algsetup{linenosize=\small}
\small
\caption{LA-DPPO for EV agents in $\mathcal{I}$}
\label{algorithm:gmappo}
\begin{algorithmic}[1]
\STATE Initialize critic and actor networks via parameters $\bm{\phi}$ and $\bm{\theta}$
\STATE Select the learning rates $\alpha^{\bm{\phi}},\alpha^{\bm{\theta}}$ of critic and actor networks
\FOR{episode (one day) $day = 1$ to $E$}
\STATE Initialize global state $s_{0}$ and local observation $o_{i,0}$
\STATE For EV agent $i \in \mathcal{I}$, establish a trajectory $\tau_{i}=[]$
\FOR{each time step (15 mins) $t = 1$ to $T$}
\FOR{EV agent $i = 1$ to $I$}
\STATE Generates action $a_{i,t}$ via the common policy $\pi_{\bm{\theta}}(a|o)$ in light of observation $o_{i,t}$
\STATE Implement agents' actions $a_{1:I,t}$ to the environment
\STATE DSO and TSO run the power flow engine and obtain LMPs and FR price signals
\STATE Observes reward $r_{i,t}$ and then the observation $o_{i,t+1}$ for next time step $t+1$
\STATE Saves this experience in trajectory $\tau_{i} \mathrel{+}= [o_{i,t},a_{i,t},r_{i,t}]$
\ENDFOR
\STATE Updates observation $o_{i,t} \leftarrow o_{i,t+1}$ for EV agent $i$
\ENDFOR
\FOR{EV agent $i = 1$ to $I$}
\STATE Estimates advantage function $\hat{\Lambda}_{i,t}$ and discounted reward-to-go $\hat{R}_{t}$ using trajectory $\tau_{i}$
\STATE Computes $\hat{\nabla}^{N,T}_{\bm{\theta}}\mathcal{L}^{a}_{i}(\bm{\theta})$ 
\IF{Satisfies the selection condition \eqref{eq:condition}}
\STATE Uploads $\ell^{a}\hat{\nabla^{a}_{i}},\ell^{c}\hat{\nabla^{c}_{i}}$
\ENDIF
\STATE The aggregator updates the parameters $\bm{\theta},\bm{\phi}$ of networks according to the rules in \eqref{eq:actor weight}-\eqref{eq:critic weight}
\ENDFOR
\ENDFOR
\end{algorithmic}
\end{algorithm}
\vspace{-0.88mm}

\section{Case Studies}
\vspace{-0.18em}
\label{sec:V}
\subsection{Experimental Setup}
\vspace{-0.00em}
\label{sec:V.A}
\subsubsection{Network and Data Descriptions}
\label{sec:V.A.1}
\begin{figure}[t!]
\centering
\subfigure[IEEE 6-bus PTN]{\includegraphics[width=0.205\textwidth]{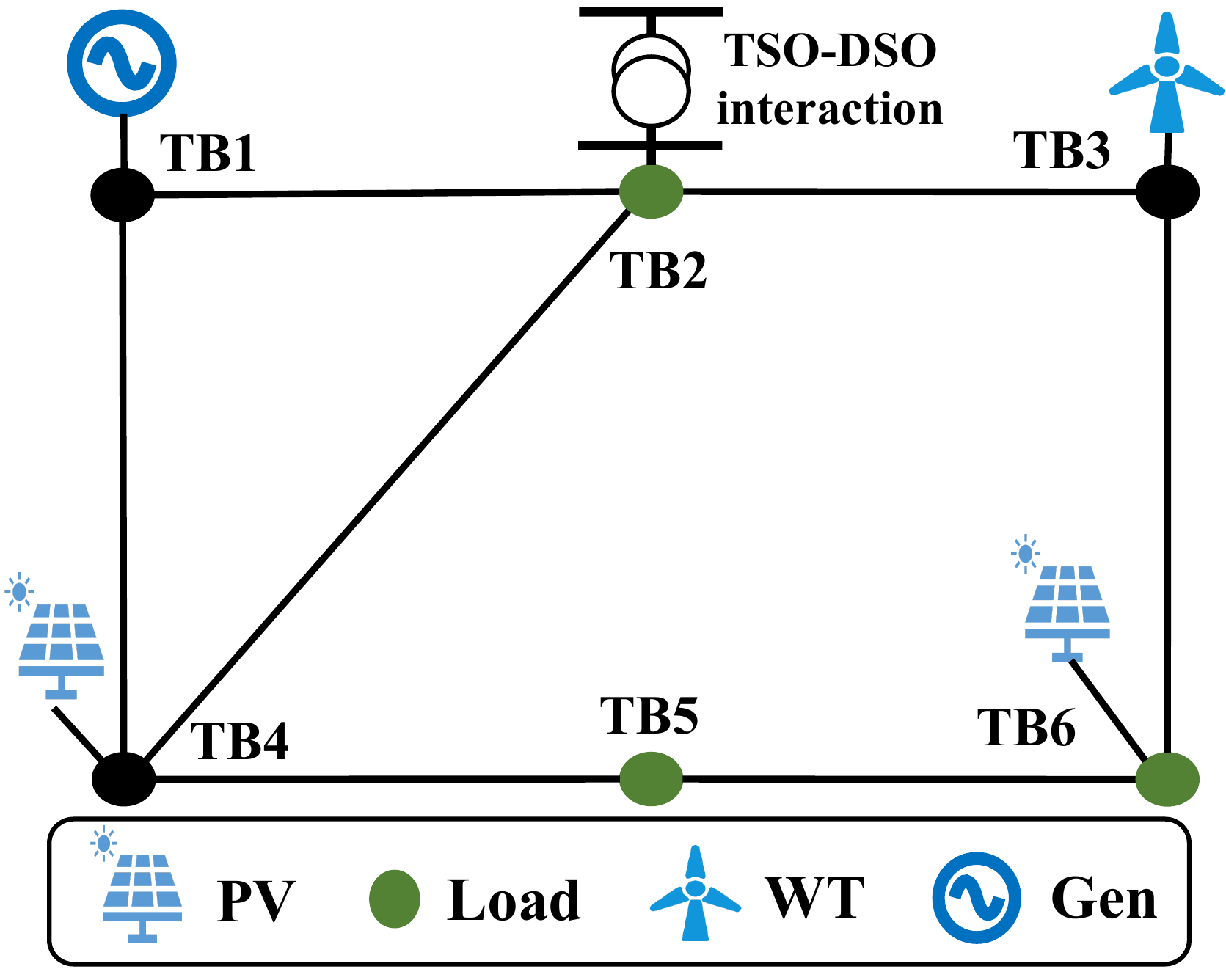}}
\subfigure[IEEE 33-bus PDN]{\includegraphics[width=0.485\textwidth]{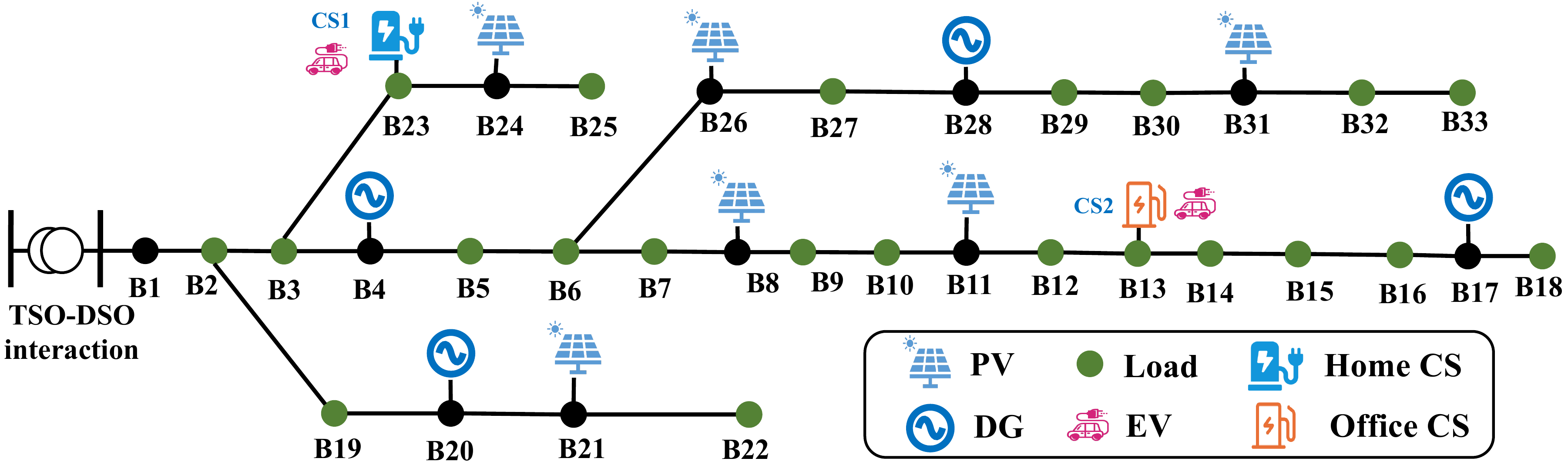}}
\vspace{-0.58em}
\caption{The studied power network including a 6-bus PTN and a 33-bus PDN.}
\label{fig:network}
\vspace{-0.38em}
\end{figure}

Case studies are firstly conducted on a power network with a modified 6-bus PTN and a modified 33-bus PDN with 120 EVs and 2 groups of CSs, e.g., office and home areas. Detailed power network structures are provided in Fig. \ref{fig:network}. To simulate the stochasticity of environment, yearly demand and RES profiles (e.g., Photovoltaics (PVs)) are gathered from Ausgrid \cite{wang2022coordinated} and subsequently divided into two sets, i.e., the training (Jan.-Nov.) and test (Dec.) sets. The truncated normal distributions that describe EV traveling uncertainties are used to initialize EV battery energy storage levels and departure time \cite{wang2022coordinated}.

\subsubsection{Benchmarks}
\label{sec:V.A.2}
The proposed LA-DPPO is compared to two RL baseline approaches and one model-based method:
i) \textbf{PPO}: every EV agent has an independent RL policy with normal policy gradients, where the original observation $o_{i,t}$ expressed in \eqref{eq:observation} is utilized as the neural network input; ii) \textbf{DPPO}: EV agents share one common RL policy under the distributed learning framework with normal policy gradients; and iii) \textbf{Distributed Control} \cite{tang2019distributed}: each group of EVs employs a distributed alternating direction method of multipliers (ADMM) method to solve a distributed coordination problem, assuming perfect information of the system models and technical parameters, handling the system uncertainties via scenario generation and reduction techniques \cite{wang2021three}.

\subsubsection{Implementations}
\label{sec:V.A.3}
The Adam optimizer is employed to update critic and actor networks in all MARL approaches with the learning rates $\alpha^{\bm{\phi}}=10^{-4}$ and $\alpha^{\bm{\theta}}=10^{-3}$. Considering the daily dynamic return of 96 time steps, the discount factor $\gamma$ and clip rate $\epsilon$ are set to 0.99 and 0.2, respectively. Two hidden layers of critic and actor networks are built using \textit{Multilayer Perceptron} (MLP) (e.g., 400 and 300 neurons). The mean and standard deviation of a Gaussian policy for actor networks is estimated via \texttt{Softplus} and \texttt{Tanh} activation functions. A total of 10,000 episodes are run with 10 random seeds, which are uniformly sampled from the training dataset. Each episode represents one operating day (e.g., 96 time steps and 24 hours).
\vspace{-0.88em}

\subsection{MARL Performance Analysis}
\begin{figure}[h!]
\vspace{-0.28em}
\centering
\subfigure[Training Reward]{\includegraphics[width=0.48\textwidth]{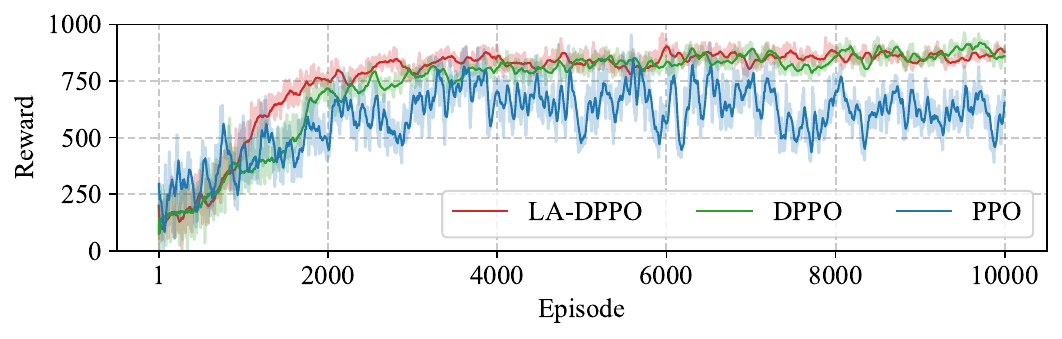}}
\subfigure[Total Communication Complexity]{\includegraphics[width=0.24\textwidth]{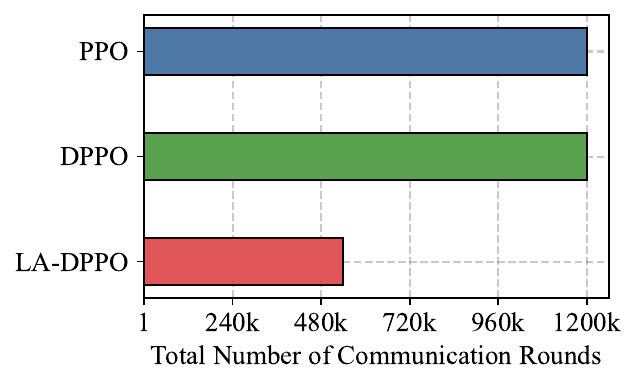}}
\subfigure[Cumulative Test Revenue]{\includegraphics[width=0.241\textwidth]{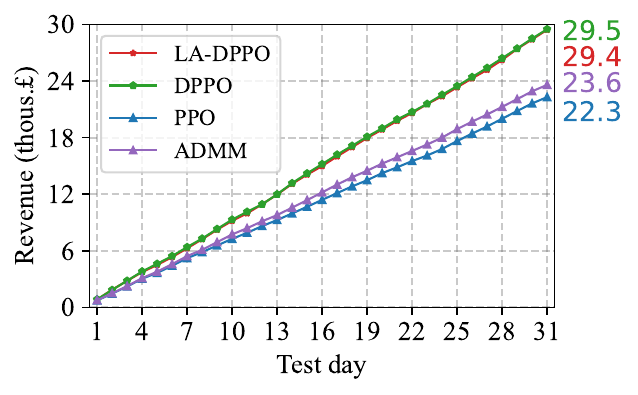}}
\vspace{-0.28em}
\caption{Episodic training reward over 10,000 episodes (a), communication complexity (b), and total test revenue over 31 days (c) for three different MARL methods and one model-based method.}
\label{fig:reward} 
\end{figure}

\vspace{-0.00em}
\label{sec:V.B}
This section analyzes the algorithm training and test performance of the suggested LA-DPPO and two MARL benchmarks (PPO and DPPO). Fig. \ref{fig:reward}(a) depicts the evolution of 120 EVs' episodic rewards across 10,000 training episodes, where the moving average across 100 episodes and the oscillations from different random seeds are respectively represented by the solid lines and shaded areas. Furthermore, we record the required communication rounds of these MARL approaches during the training process in Fig. \ref{fig:reward}(b) and illustrate the cumulative revenue of 120 EVs over the test month in Fig. \ref{fig:reward}(c) after implementing the trained RL (actor) policy.

It can be observed from Fig. \ref{fig:reward}(a) that LA-DPPO (red) and DPPO (green) converge to the same reward level around 938. However, by deploying the LAPG, the communication complexity of training LA-DPPO has been significantly reduced (over 50\% communication rounds have been skipped as depicted in Fig. \ref{fig:reward}(b)), compared with DPPO based on normal policy gradient. In detail, considering that there are 10,000 recorded iterations during the training stage, the communication rounds of information change for both PPO and DPPO are 1200k, i.e., there is information exchange for each agent in each iteration, while the proposed LA-DPPO only requires around 540k communication rounds of information change to finish the MARL training. It can be anticipated that the reduction can be even more significant when a larger number of agents are considered, which can reduce the communication overhead and save massive amounts of communication resources. On the other hand, it can be observed that the training reward and communication complexity of PPO (blue) both exhibit a worse performance than the proposed LA-DPPO. This can be expected since LA-DPPO employs the distributed framework based on the parameter sharing scheme for the parallel learning of a common policy instead of personalizing RL policies for every EV agent, which can reduce training time, address the non-stationarity issue, and stabilize training performance. Overall, it can be concluded that LA-DPPO is capable of achieving communication-efficient and stable MARL training without performance degradation.

Similar to the training performance, as illustrated in Fig. \ref{fig:reward}(c), the suggested LA-DPPO during the test process achieves almost the same performance as the DPPO, while significantly outperforming ADMM by 24.6\% and PPO by 31.8\% the cumulative revenue level over 31 test days. The reason is that the suggested distributed learning framework based on parameter sharing can better stabilize the algorithm performance and deal with uncertainties associated with RESs and demand. Furthermore, after training, all three MARL approaches (PPO, DPPO, and LA-DPPO) can be deployed in real-time within 1 sec., which means EV agents can independently observe environment information and make scheduling decisions without interactions with other agents. However, the model-based ADMM method requires around 413.34 sec. to solve the optimization iteratively; thus, there will be communications and information exchange between EV agents for each iteration. To summarize, the advantage of the proposed model-free learning method for timely decision-making has been exhibited thoroughly in both the training and test stages.
\vspace{-0.7em}

\subsection{EV Dispatching Analysis}
\begin{figure*}[t!]
\centering
\includegraphics[width=1\textwidth]{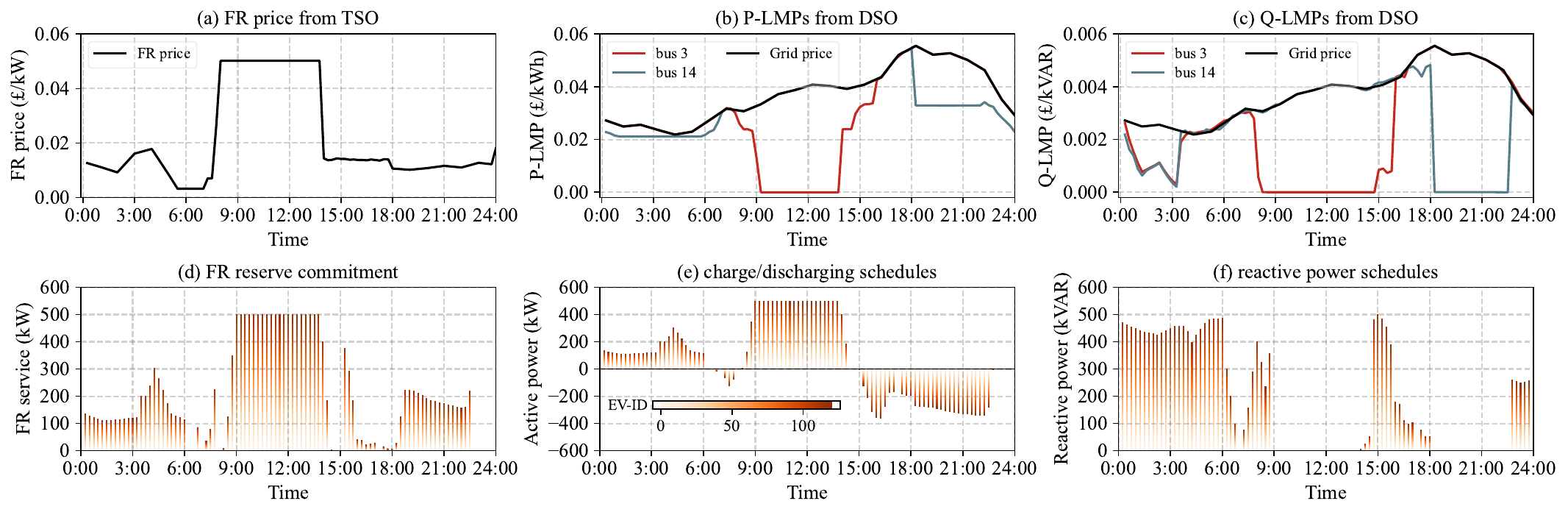}
\vspace{-1.08em}
\caption{Shadow price signals and aggregated EV scheduling behaviors: a) FR price signals from TSO, b) P-LMPs from DSO, c) Q-LMPs from DSO, d) FR reserve commitment, e) active power charging/discharging, and f) reactive power scheduling.}
\label{fig:ev_schedule} 
\end{figure*}

\begin{figure}[t!]
\vspace{-0.18em}
\centering
\includegraphics[width=0.465\textwidth]{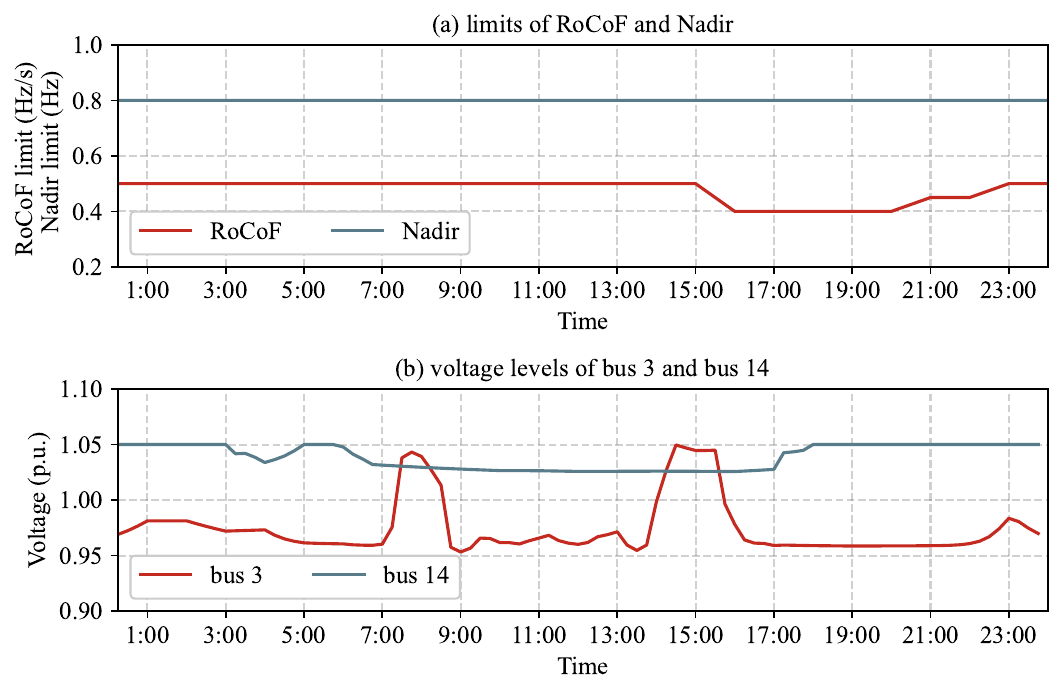}
\vspace{-1.08em}
\caption{Frequency and voltage profiles.}
\label{fig:f_v_profile} 
\vspace{-0.48em}
\end{figure}

\vspace{-0.0em}
\label{sec:V.C}
The purpose of this section is to verify the well-trained policy of LA-DPPO for EV dispatching characteristics in PDN, as shown in Fig. \ref{fig:ev_schedule}. Note that the FR price in Fig. \ref{fig:ev_schedule}(a) and the grid price in Fig. \ref{fig:ev_schedule}(b) are obtained from day-ahead TSO-DSO interaction at stage 1. Furthermore, to estimate the contribution of EVs to TSO-FR service and DSO-VR support, related frequency and voltage profiles are presented in Fig. \ref{fig:f_v_profile}.

\subsubsection{EV Service Provisions}
In response to FR price signals and LMPs in Fig. \ref{fig:ev_schedule}(a)-(c), 120 EVs are making FR and active/reactive power dispatching behaviors, as depicted in Fig. \ref{fig:ev_schedule}(d)-(f). Specifically, EVs deliver FR service in response to the FR price signals shown in Fig. \ref{fig:ev_schedule}(a). The FR price is at its highest level during the midday due to the low-inertia level caused by high PV generation and potential RES curtailment, while the FR price is at its lowest level in the early morning and evening because of the high net demand level. Accordingly, EVs provide much more FR service in the midday, while providing less service in the evening. It can be observed from Fig. \ref{fig:f_v_profile}(a) that frequency Nadir (blue) and RoCoF (red) are respectively limited to be higher than 49.2 Hz and lower than 0.5 Hz/s, maintaining frequency security.

Fig. \ref{fig:ev_schedule}(e) illustrates that most EVs decide to charge power in the morning and midday, while discharging power in the evening. It is evident from Fig. \ref{fig:ev_schedule}(b) that the P-LMPs at bus 14 (blue) in the morning (0-7 hours) and the evening (18-24 hours) are significantly lower than the grid price signals (black), meaning EVs power discharging behaviors are less encouraged. The price differentials discussed above can also be reflected in the voltage level at bus 14 (blue), as depicted in Fig. \ref{fig:f_v_profile}(b), which has almost reached the maximum limit at 1.05 p.u. during these two time periods. The potential reasons are the CS located on bus 13 and the conventional DG (e.g., diesel generator) located on bus 17, contributing to the high voltage levels on bus 14. Accordingly, in response to the P-LMPs, EVs decide to charge power in the early morning and discharge less power in the evening for voltage support, while charging more power at bus 3 (red) in the midday due to the zero P-LMPs (Fig. \ref{fig:ev_schedule}(b)) caused by the high PV generation. The midday charging behaviors also mitigate the voltage rise issues at bus 3 (red) induced by high PV generation, as depicted in Fig. \ref{fig:f_v_profile}(b). 

Similarly, due to the high voltage level at bus 14 (blue) shown in Fig. \ref{fig:f_v_profile}(b), the Q-LMPs (blue) are much lower than the grid price signals (black), especially in the evening (reaching zero), as depicted in Fig. \ref{fig:ev_schedule}(c). Accordingly, EVs decide to provide reactive power in the early morning rather than the evening, because of the zero Q-LMPs for voltage support.

\subsubsection{EV Revenues}
Table \ref{table:revenue_big} summarizes that 120 EVs can obtain considerable revenues through reasonable active/reactive power and FR dispatching behaviors in response to P-LMPs, Q-LMPs, and FR price signals. 

\subsection{Scalability Analysis}
\label{sec:V.D}
\begin{figure*}[t!]
\centering
\includegraphics[width=1\textwidth]{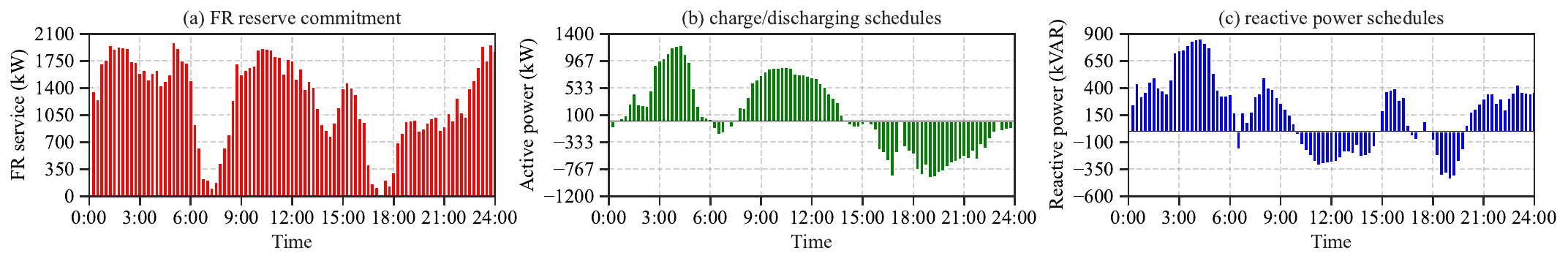}
\vspace{-1.68em}
\caption{Aggregated EV scheduling behaviors in the large power system: a) FR reserve commitment, b) active power charging/discharging, and c) reactive power scheduling.}
\label{fig:big_sum} 
\end{figure*}

\begin{figure}[t!]
\centering
\includegraphics[width=0.475\textwidth]{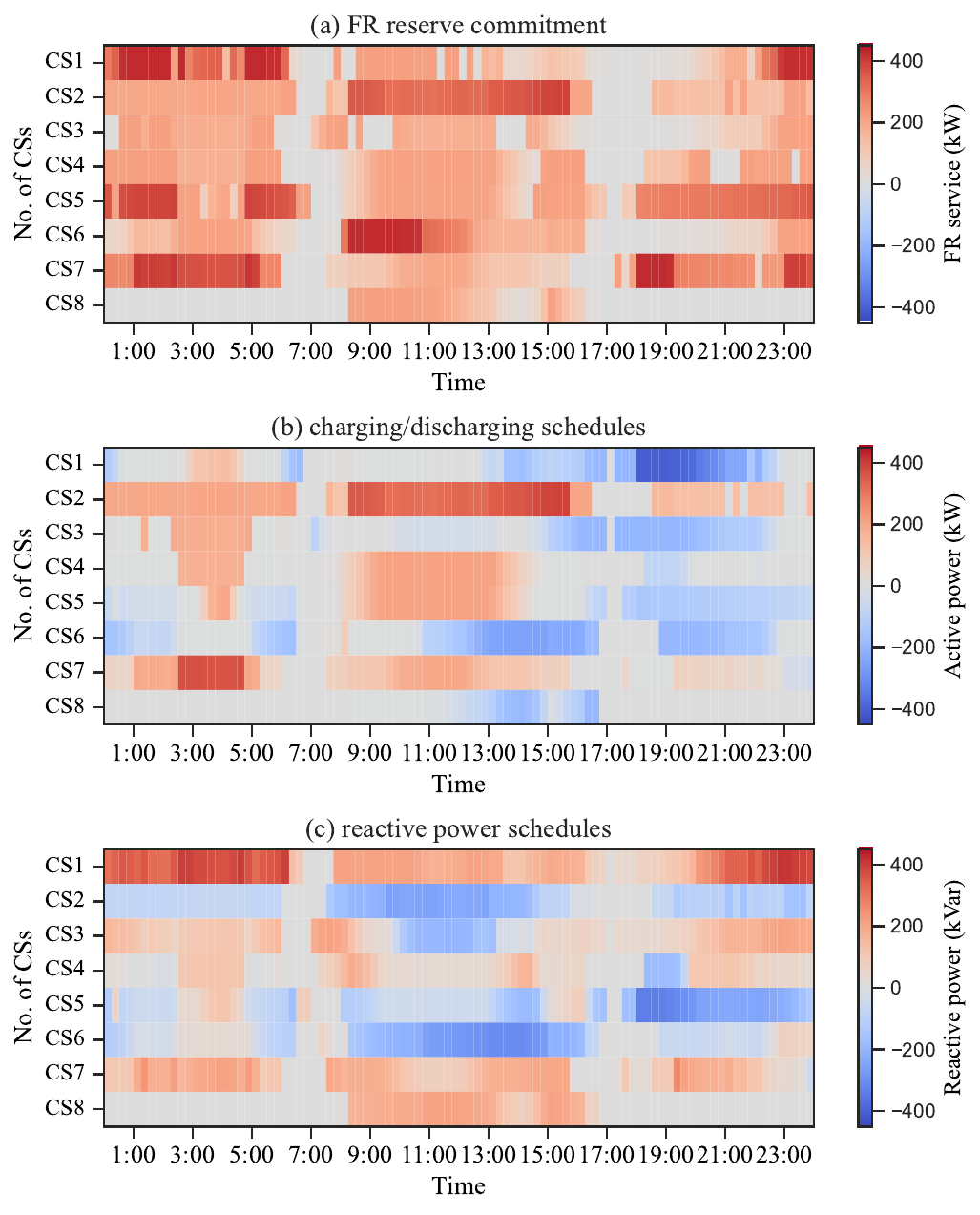}
\vspace{-0.48em}
\caption{Aggregated EV dispatching behaviors in the CSs of the larger power network: a) FR reserve commitment, b) active power charging/discharging, and c) reactive power scheduling.}
\label{fig:big_cs} 
\vspace{-0.18em}
\end{figure}

This section aims to demonstrate the scalability of the suggested LA-DPPO approach for the coordinated TSO-FR and DSO-VR service provision problem of multi-EVs, where a modified 69-bus power network with 8 groups of CSs and 500 EVs is deployed. The test performances of 500 EVs in FR reserve commitment, active power charging/discharging, and reactive power scheduling are illustrated in Fig. \ref{fig:big_sum} and Fig. \ref{fig:big_cs}.

It can be observed from Fig. \ref{fig:big_sum} that EVs are connected with CSs to provide FR service and make active and reactive power scheduling behaviors in response to LMP signals. Specifically, Fig. \ref{fig:big_sum} (a) and (b) show that most EVs prefer to provide FR service and charge power in the morning and midday, while discharging power in the evening, because of the low P-LMPs in the midday and the high P-LMPs in the evening. It can also be found that the time periods of FR service provision and charging behaviors are highly overlapping, which is aligned with the conclusion that EVs can effectively provide a higher volume of FR service via a swift switch from charging mode to discharging mode when they are charging power. Furthermore, as illustrated in Fig. \ref{fig:big_sum}(c), EVs prefer to absorb reactive power in the midday while injecting reactive power in the morning towards effective voltage regulations. The detailed EV schedules in all the CSs are illustrated in Fig. \ref{fig:big_cs}, which demonstrates a similar conclusion to Fig. \ref{fig:big_sum}. For example, EVs at CS 2 provide FR reserve and absorb reactive power from the grid almost all day, especially in the midday when PV penetration is high. Most EVs are on the road back home during 16:30-18:30, while EVs at CS 6 choose to discharge power during 18:30-21:00; thus, there is only little FR reserve support from CS 6 during 16:30-21:00. This phenomenon can also be observed in CS 1 and CS 3. In addition, it can be found that there are almost no EV schedules at CS 8 in the early morning and evening. This is because CS 8 only represents a working area, where EVs mainly choose to connect in the late morning and afternoon. Note that detailed EV schedules at different CSs may vary, depending on the specific system and CS settings. Overall, the effectiveness and scalability of the suggested LA-DPPO approach in coordinating large-scale EVs for service provisions have been verified.

Table \ref{table:revenue_big} demonstrates that the total revenues of providing FR service and VR support increase, when the EV number increases from 500 to 700. Additionally, the revenues from active and reactive power support may not rise proportionally as the EV number increases, due to the potential limitations of DSO flexibility and CS capacity. However, it can still be concluded that both TSO and DSO have benefited more from the increase in EV numbers, further verifying the scalability of the proposed LA-DPPO approach in learning effective control policies for different EV numbers.
\vspace{-0.48em}

\begin{table}[h!]
\footnotesize
\centering
\renewcommand\arraystretch{1.00}
\setlength{\abovecaptionskip}{3pt}
\caption{Scalability analysis for different EV numbers.}
\setlength{\tabcolsep}{1.68mm}{
\begin{tabular}{ |c|c|c|c|c| }
 	  \toprule
        \begin{tabular}[c]{@{}c@{}}No. \\of EVs \end{tabular}
         & \begin{tabular}[c]{@{}c@{}} FR service\\revenue (\pounds)\end{tabular}
 	  & \begin{tabular}[c]{@{}c@{}} Active power\\revenue (\pounds)\end{tabular}
 	  & \begin{tabular}[c]{@{}c@{}} Reactive power\\revenue (\pounds)\end{tabular}
        & \begin{tabular}[c]{@{}c@{}} Total \\revenue (\pounds)\end{tabular} \\ \midrule 
        120  & 728.4 & 200.9 & 31.6 & 960.9\\  \midrule 
        500  & 2980.8 & 841.0 & 139.5 & 3961.3\\ \midrule 
        600  & 3442.0 & 967.2 & 142.8 & 4552.0\\ \midrule 
        700  & 3837.2 & 1053.0 & 148.4 & 5038.6\\ \bottomrule      
\end{tabular}}
\label{table:revenue_big}
\end{table}

\subsection{Performance Verification via Dynamic Simulations}
\begin{figure}[h!]
\centering
\includegraphics[width=0.46\textwidth]{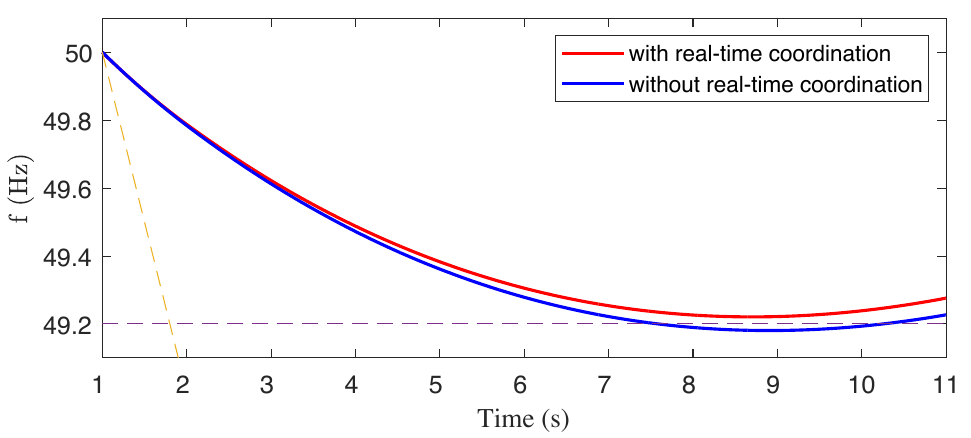}
\vspace{-0.48em}
\caption{Comparison of frequency dynamic evolution curves after a power infeed loss in the transmission network: a) with real-time TSO-FR and DSO-VR coordination of EVs, b) without real-time TSO-FR and DSO-VR coordination of EVs.}
\label{fig:big_freq} 
\end{figure}

\vspace{-0.08em}
\label{sec:V.E}
To further verify the performance of the proposed two-stage service provision framework in ensuring the frequency security of the transmission network, one fault scenario at the transmission level (e.g., 500 MW power infeed loss) has been simulated to demonstrate the FR service provision behaviors of EVs and corresponding frequency evolution curve involving both frequency Nadir and RoCoF. Specifically, a comparison is conducted, including two cases: a) the proposed two-stage service provision framework is adopted, while both TSO-FR and DSO-VR are embedded into the reward function to promote EVs to provide real-time frequency and voltage services; b) only DSO-VR is embedded into the reward function, while the real-time TSO-FR commitment of EVs at stage 2 is not considered. Detailed comparison results between the above two cases are depicted in Fig. \ref{fig:big_freq}.

It can be observed from Fig. \ref{fig:big_freq} that frequency security has been ensured under the case `with real-time coordination of TSO-FR and DSO-VR', i.e., maintaining frequency Nadir above 49.2 Hz and RoCoF is lower than 1 Hz/s. In detail, a total of 450 MW FR from conventional generators and 56 MW FR from EVs have been enabled to support post-fault frequency security. However, when the real-time coordination of TSO-FR and DSO-VR is not considered, some EVs may not be able to provide TSO-FR due to the lack of incentive schemes and the potential influence of uncertainties, which eventually leads to the reduction of available FR from EVs (e.g., only 44 MW FR from EVs is activated for post-fault frequency security). Consequently, as shown in Fig. \ref{fig:big_freq}, there is a certain level of violation of the frequency Nadir constraint, i.e., frequency Nadir reaches 49.18 Hz, which is less than 49.2 Hz and may trigger insecure system operations. In this context, the advantages of the proposed two-stage service provision framework in ensuring post-fault frequency security at real-time stage have been demonstrated thoroughly.

Furthermore, it would be interesting to discuss the performance of the proposed framework and the learning algorithm after the potential fault. From the perspective of TSO, when a fault in the transmission level happens, EVs can promptly use the available FR reserve to participate in TSO frequency support. From the perspective of DSO, if the fault in the transmission level introduces certain changes in the power flow and voltage levels at the distribution network, this will be directly reflected by DSO P-LMPs and Q-LMPs. In other words, when the system condition changes (e.g., a fault occurs), LMPs will also change and then guide EVs to make realistic active and reactive power dispatches in response to the current system condition. Due to the effective incentive scheme based on shadow price signals, the proposed learning-based algorithm can be flexible enough to guide EVs to make realistic dispatches under different system conditions. In addition, due to the multi-agent setting and the decentralized framework, well-trained RL policies can guide EV agents to make independent decisions in a computationally efficient manner without interactions with other agents.

\section{Conclusions}
\label{sec:VI}
This paper proposes a two-stage service provision framework for multi-EVs in the context of TSO-DSO interactions towards coordinated TSO-FR and DSO-VR support. Day-ahead TSO-DSO interaction is formulated at the first stage for FR reserve schedules, while real-time EV dispatching behaviors are realized in a decentralized POMG at the second stage to deliver the scheduled FR service while considering DSO-VR support via proactive active and reactive power dispatching. A new MARL method called LA-DPPO is proposed to handle the POMG at the second stage, which is distinguished by a distributed learning framework with LAPG update and parallel learning for communication-efficient and stable MARL training. Experiments are conducted on a power network with a 6-bus PTN and a 33-bus PDN as well as a 69-bus PDN to assess the superiority of the suggested MARL approach in optimality, communication efficiency, and stability, in comparison with existing cutting-edge MARL methods. Furthermore, the decision-making of multi-EVs on coordinated TSO-FR and DSO-VR provisions is analyzed, while the scalability of the proposed method is verified.

Furthermore, future work aims to enhance the studied problem from two directions. First, this paper aims to investigate the coordinated service provision problem of large-scale EVs in the context of TSO-DSO interactions. The information barrier and potential privacy concerns between transmission and distribution should be better considered in the proposed two-stage framework. Therefore, future work will try to develop more realistic privacy-preserving mechanisms for day-ahead TSO-DSO interactions, taking both operational efficiency and regulatory compliance into account. Second, this work mainly focuses on the perspective of normal operation and frequency-constrained scheduling, e.g., how to prepare sufficient reserve for the potential fault. Therefore, it will be interesting to develop effective corrective control schemes for multi-EVs, which can guide EVs to take corrective actions against various emergency situations towards more reliable and resilient system operations.

\bibliographystyle{IEEEtran}
\bibliography{References.bib}
\ifCLASSOPTIONcaptionsoff
  \newpage
\fi

\begin{IEEEbiography}[{\includegraphics[width=1in,height=1.25in,clip,keepaspectratio]{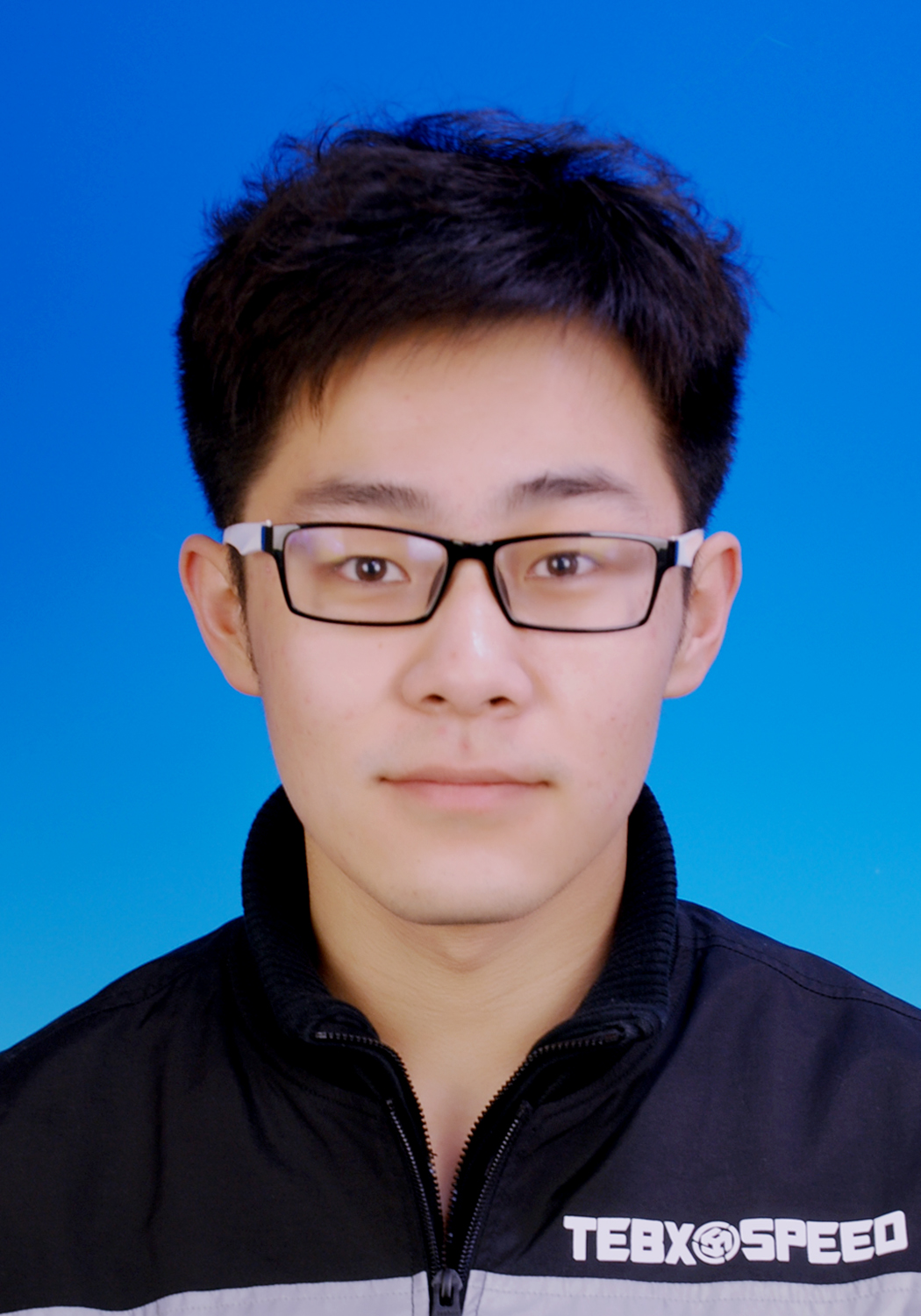}}]{Yi Wang}
received the B.Eng. degree and the M.Eng. degree from Tianjin University in 2015 and 2018, and the Ph.D. degree from Imperial College London in 2022. He is currently employed as a Research Associate in the Department of Electrical and Electronic Engineering at Imperial College London. His research interests include mathematical programming and learning approaches applied to the planning and operation of networked microgrids, the resilience enhancement of future power systems, frequency-constrained power system optimisation, and multi-energy system integration.
\end{IEEEbiography}

\begin{IEEEbiography}[{\includegraphics[width=1in,height=1.25in,clip,keepaspectratio]{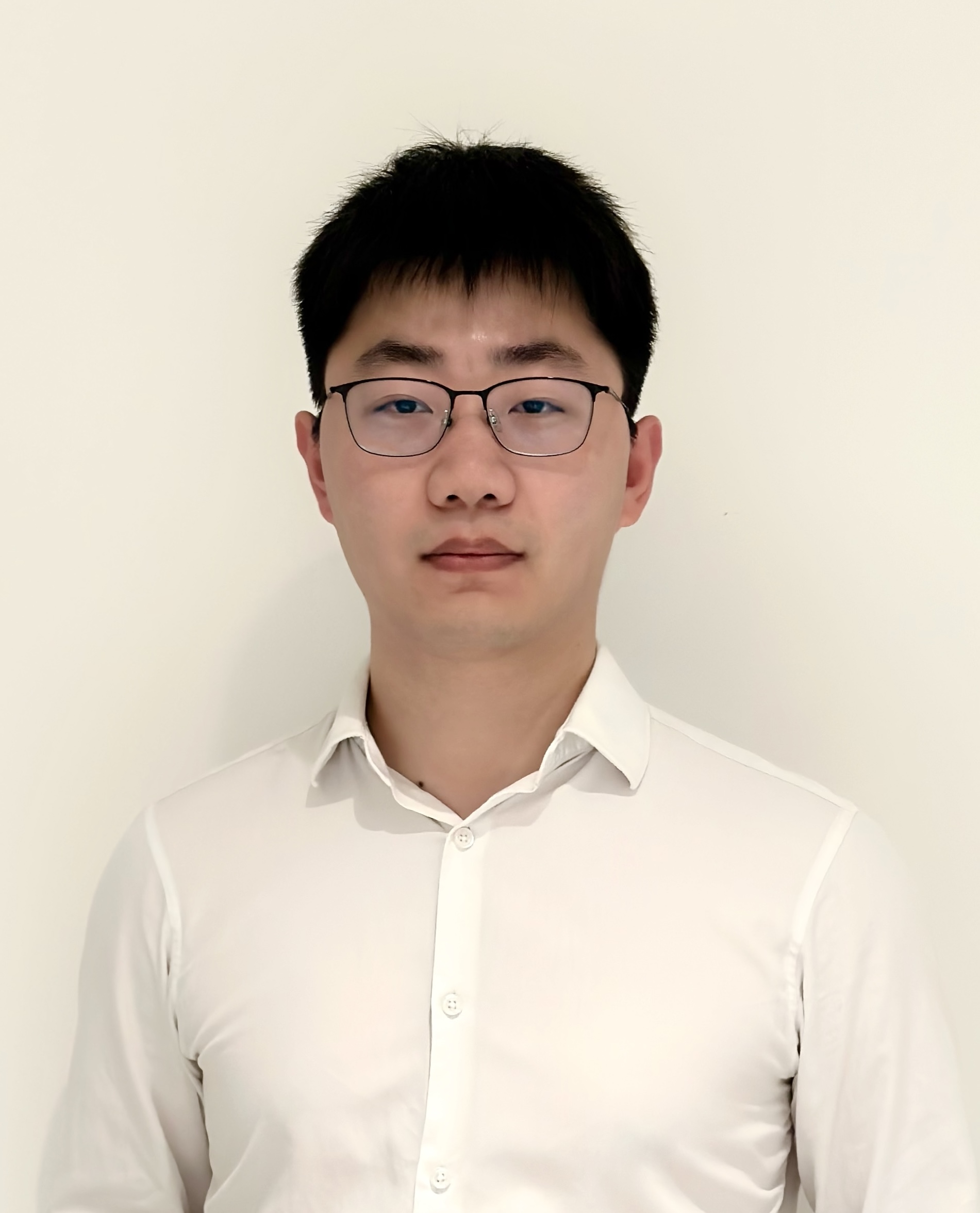}}]{Dawei Qiu}
received the B.Eng. degree from Northumbria University in 2014, the M.Sc. degree from University College London in 2015, and the Ph.D. degree from Imperial College London in 2020. He was employed as a Research Associate in the Department of Electrical and Electronic Engineering at Imperial College London, then promoted to be a Research Fellow in Market Design for Low-Carbon Energy Systems in 2023. He is currently a Lecturer at Smart Energy System in the Department of Engineering at the University of Exeter. His research focuses on the development and application of data-driven AI approaches to electricity market, peer-to-peer energy trading, multi-energy system, microgrid resilience, and vehicle-to-grid flexibility.
\end{IEEEbiography}

\begin{IEEEbiography}[{\includegraphics[width=1in,height=1.25in,clip,keepaspectratio]{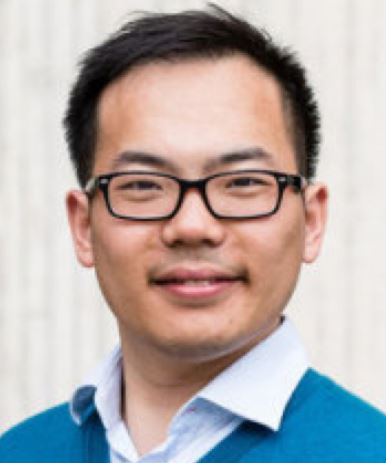}}]{Fei Teng}
received the B.Eng. degree in electrical engineering from Beihang University, China, in 2009, and the M.Sc. and Ph.D. degrees in electrical engineering from Imperial College London, U.K., in 2010 and 2015, respectively, where he is currently a Senior Lecturer with the Department of Electrical and Electronic Engineering. His research focuses on the power system operation with high penetration of Inverter-Based Resources (IBRs) and the Cyber-resilient and Privacy-preserving cyber-physical power grid.
\end{IEEEbiography}

\begin{IEEEbiography}[{\includegraphics[width=1in,height=1.25in,clip,keepaspectratio]{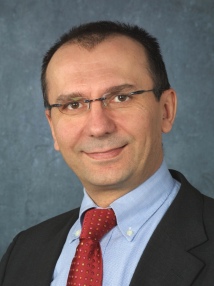}}]{Goran Strbac}
is a Professor of Energy Systems at Imperial College London, London, U.K. He led the development of novel advanced analysis approaches and methodologies that have been extensively used to inform industry, governments, and regulatory bodies about the role and value of emerging new technologies and systems in supporting cost effective evolution to smart low carbon future. He is currently the Director of the joint Imperial-Tsinghua Research Centre on Intelligent Power and Energy Systems, Leading Author in IPCC WG 3, Member of the European Technology and Innovation Platform for Smart Networks for the Energy Transition, and Member of the Joint EU Programme in Energy Systems Integration of the European Energy Research Alliance.
\end{IEEEbiography}

\end{document}